\documentclass[12pt, letterpaper]{article}
\usepackage[utf8]{inputenc}
\usepackage[margin=1in]{geometry}
\usepackage{setspace}
\usepackage{graphicx}
\usepackage{booktabs}
\usepackage{amsmath}%
\usepackage[labelfont=bf, skip=5pt]{caption}
\usepackage[style=chem-acs, articletitle=true]{biblatex}
\usepackage [english]{babel}
\usepackage [autostyle, english = american]{csquotes}
\MakeOuterQuote{"}

\begin{document}

\begin{center}
    \huge{A Low Temperature Structural Transition in Canfieldite, Ag$_8$SnS$_6$, Single Crystals} \\[10pt]
    \vskip 0.25cm
    \large{Tyler J. Slade$^{1,2}$, Volodymyr Gvozdetskyi$^3$, John M. Wilde$^{1,2}$, Andreas Kreyssig$^{1,2}$, Elena Gati$^{1,2}$, Lin-Lin Wang$^{1,2}$, Yaroslav Mudryk$^{1}$, Raquel A. Ribeiro$^{1,2}$, Vitalij K. Pecharsky$^{1,4}$, Julia V. Zaikina$^3$, Sergey L. Bud’ko$^{1,2}$, Paul C. Canfield$^{1,2}$} \\[3pt]
\end{center}

\textit{$^{1}$Ames Laboratory, US DOE, Iowa State University, Ames, Iowa 50011, USA} \\
\textit{$^{2}$Department of Physics and Astronomy, Iowa State University, Ames, Iowa 50011, USA}
\\
\textit{$^{3}$Department of Chemistry, Iowa State University, Ames, Iowa 50011, USA}
\\
\textit{$^{4}$Department of Materials Science and Engineering, Iowa State University, Ames, IA 50011, USA}

\begin{abstract}

\noindent Canfieldite, Ag$_8$SnS$_6$, is a semiconducting mineral notable for its high ionic conductivity, photosensitivity, and low thermal conductivity. We report the solution growth of large single crystals of Ag$_8$SnS$_6$ of mass up to 1 g from a ternary Ag-Sn-S melt.  On cooling from high temperature, Ag$_8$SnS$_6$ undergoes a known cubic (\textit{F}$\bar{4}3$\textit{m}) to orthorhombic (\textit{Pna}2$_1$) phase transition at $\approx$ 460 K. By studying the magnetization and thermal expansion between 5–300 K, we discover a second structural transition at $\approx$ 120 K.  Single crystal X-ray diffraction reveals the low temperature phase adopts a different orthorhombic structure with space group \textit{Pmn}2$_1$ (\textit{a} = 7.6629(5) \AA, \textit{b} = 7.5396(5) \AA, \textit{c} = 10.6300(5) \AA\, \textit{Z} = 2 at 90 K) that is isostructural to the room temperature forms of the related Se-based compounds Ag$_8$SnSe$_6$ and Ag$_8$GeSe$_6$. The 120 K transition is first-order and has a large thermal hysteresis. Based on magnetization and thermal expansion data, the room temperature polymorph can be kinetically arrested into a metastable state by rapidly cooling to temperatures below 40 K. We lastly compare the room and low temperature forms of Ag$_8$SnS$_6$ with its argyrodite analogues, Ag$_8$\textit{TQ}$_6$ (\textit{T} = Si, Ge, Sn; \textit{Q} = S, Se), and identify a trend relating the preferred structures to the unit cell volume, suggesting smaller phase volume favors the \textit{Pna}2$_1$ arrangement. We support this picture by showing that the transition to the \textit{Pmn}2$_1$ phase is avoided in Ge alloyed Ag$_8$Sn$_{1-x}$Ge$_x$S$_6$ samples as well as pure Ag$_8$GeS$_6$.

\end{abstract}

\newpage

\section{Introduction}

\begin{doublespace}

Multinary silver chalcogenides exhibit rich structural chemistry and electronic properties that generate both fundamental and technological interest. A hallmark of these compounds is weak bonding between silver and chalcogenide atoms, often resulting in diverse crystal structures, substantial anharmonic atomic displacements, site occupancy disorder, and/or metal atoms in unique, distorted coordination environments.\autocite{wiegers1971crystal,kuhs1979argyrodites,wood1992synthesis,li1995csag5te3}  Consequentially, many silver chalcogenides exhibit high ionic conductivity and a tendency to undergo phase transitions upon heating or cooling.\autocite{yakshibayev1989ionic,junod1977metal,boyce1979superionic,rettie2018ag2se} The soft bonding and structural complexity can give rise to low phonon velocities and intrinsically glasslike thermal conductivity. As such, many silver chalcogenides are attractive thermoelectric materials.\autocite{morelli2008intrinsically,roychowdhury2021enhanced,lin2016concerted,kim2005crystal,charoenphakdee2009ag8site6,acharya2018enhancement,lin2021thermally}  Compared to oxides, the bonds in metal chalcogenides are generally more covalent, which affords smaller band gaps and higher charge carrier mobility,\autocite{zeier2016thinking} making some chalcogenides promising for solar energy conversion.\autocite{stroyuk2018solar}

The past decades witnessed impressive advancement in the synthesis of chalcogenide materials.  Whereas the high vapor pressures and low boiling points of the chalcogens, particularly sulfur, are challenges for the growth of large single crystals of chalcogen-rich compounds, solution or flux techniques have proven invaluable for their ability to permit chemical reactions and crystallization at moderate temperatures compared to traditional solid state methods.\autocite{lin2012development,kanatzidis2017discovery,kanatzidis1997new} In this regard, polychaclcogenide salts are powerful solvents for the synthesis of alkali metal and main group chalcogenide compounds.\autocite{kanatzidis1997new,kanatzidis1990molten} Low-melting binary metal-sulfur compositions or eutectics can also be effective starting points for solution growth of sulfur rich materials.\autocite{lin2012development}

Canfieldite,\autocite{palache1927memorial} Ag$_8$SnS$_6$, is a mineral belonging to the family of closely related argyrodite compounds.\autocite{kuhs1979argyrodites} Like all argyrodites, Ag$_8$SnS$_6$ adopts a high temperature cubic crystal structure with the \textit{F}$\bar{4}3$\textit{m} space group that features a highly disordered Ag sublattice in which the Ag ions are distributed over 3 partially occupied crystallographic sites. Previous work indicates that at 460 K, Ag$_8$SnS$_6$ undergoes a transition from the high temperature (ht) cubic phase to the room temperature (rt) orthorhombic structure with space group \textit{Pna}2$_1$.\autocite{kuhs1979argyrodites,mikolaichuk2010phase,shen2020high} Upon heating, the structural transition can be visualized as the Ag sublattice “melting” into the highly disordered state while the Sn and S framework remains essentially rigid, or intact. As a consequence, the ht phase has high ionic conductivity supported by mobile Ag$^+$ ions.  The heavy disorder in the “liquid like” Ag sublattice facilitates exceptionally strong phonon scattering, and the lattice thermal conductivity of ht-Ag$_8$SnS$_6$ approaches the theoretical amorphous minimum.\autocite{shen2020high} Accordingly, recent work demonstrates that ht-Ag$_8$SnS$_6$ is a promising thermoelectric material with a figure of merit \textit{zT} of 0.8 near 773 K when alloyed with Se.\autocite{shen2020high,ghrib2015high} Most work on the rt, orthorhombic, variant is concerned with nanocrystals,\autocite{an2003synthesis,li2000synthesis,li2001preparation} which have been prepared and studied as counter electrodes in solar cells and for photocatalytic water splitting and dye degradation.\autocite{he2015role,cheng2016photo,shambharkar2016ethylene} Notably, the rt phase has a band gap of approximately 1.4 eV and high absorption coefficient of 104 cm$^{-1}$ in the visible range,\autocite{chun2015electronic} making rt Ag$_8$SnS$_6$ a potential light absorber for photovoltaic applications.\autocite{zhu2018application,boon2018ag} 

Whereas both the ambient and high temperature forms of Ag$_8$SnS$_6$ exhibit interesting properties and structural chemistry, the evolution of the crystal structure and physical properties below room temperature are yet to be reported. Here, we demonstrate solution growth of large ($\approx$1 g) single crystals of Ag$_8$SnS$_6$ out of a Sn-S rich ternary melt, and by use of temperature dependent magnetization, thermal expansion, and X-ray diffraction measurements, we identify a previously unreported first-order structural phase transition at 120 K. We find that below 120 K, Ag$_8$SnS$_6$ can adopt a different low temperature (lt) orthorhombic structure with the \textit{Pmn}2$_1$ space group that is isostructural to the room temperature forms of the related Se-based compounds Ag$_8$SnSe$_6$ and Ag$_8$GeSe$_6$. The rt to lt structural phase transition has large thermal hysteresis, and the magnetization and thermal expansion results indicate the room temperature phase can be kinetically arrested by rapid cooling into a metastable state below 40 K. By comparing the room and low temperature Ag$_8$SnS$_6$ polymorphs with the preferred structures of the argyrodite analogues Ag$_8$\textit{TQ}$_6$ (\textit{T} = Si, Ge, Sn; \textit{Q} = S, Se), we suggest that lower volume unit cells (normalized per formula unit) favor the rt \textit{Pna}2$_1$ form. We support this picture by applying chemical pressure, i.e. reducing the unit cell volume, to Ag$_8$SnS$_6$ by partially replacing Sn with Ge, and find the rt \textit{Pna}2$_1$ phase is preserved down to 5 K in Ag$_8$Sn$_{1-x}$Ge$_x$S$_6$ and pure Ag$_8$GeS$_6$. Lastly, we use density functional theory calculations to discuss the role of Ag-S bonding in dictating the energetically preferred structure.

\section{Experimental Details}

\textbf{\textit{Crystal growth:}} Both of the Ag-S and Sn-S binary phase diagrams have accessible liquid regions for compositions near Ag$_2$S and compositions between SnS$_2$ and SnS with eutectic temperatures as low as 740°C.\autocite{sharma1986ag,sharma1986s,okamoto2000phase}  Perhaps more importantly, studies of parts of the ternary Ag-Sn-S phase diagram indicate the existence of an exposed liquidus surface for the primary formation of Ag$_8$SnS$_6$.\autocite{wang1989experimental}  Based on the existing phase diagram data, as well as our experience with S-based growths, we used Ag$_2$S (Alfa-Aesar, 99.5 $\%$), S (Alfa-Aesar 99.99$\%$) and Sn (Alfa-Aesar, 99.99$\%$) in a molar ratio 35:38:27 respectively, giving a stoichiometry of Ag$_{41}$Sn$_{16}$S$_{43}$ as the initial composition of our melt.

To test our ability to safely grow Ag$_8$SnS$_6$ out of the ternary melt in a well controlled manner, we first placed the Ag$_2$S-S-Sn mixture into a fritted alumina crucible set,\autocite{canfield2016use} sealed the crucible set into an amorphous silica ampule under a 1/4 atm of high purity Ar, and gradually heated to 800°C over 10 hours.  After holding at 800°C for 10 hours, the ampule was removed from the furnace, inverted into a centrifuge, and decanted.\autocite{canfield2016use,canfield2019new} All of the materials passed through the frit with (i) no apparent attack on the alumina crucible, (ii) no evaporative loss of material from the crucible, (iii) no evidence for over-pressure in the growth ampule.  Based on these results, we performed a second growth, again heating to 800°C, dwelling for 10 h, and then cooling over 200 hours to 625°C.  After decanting the excess liquid, large, mirror-faceted, single crystals of Ag$_8$SnS$_6$ weighing up to 1 gram with dimensions as large as 5-10 mm were recovered (see the inset of Figure 1).  Ge alloyed samples were prepared in the same manner, substituting elemental Ge (Alfa-Aesar, 99.99$\%$) for Sn using initial compositions of Ag$_{41}$(Sn$_{1-x}$Ge$_x$)$_{16}$S$_{43}$ to produce single crystals of Ag$_8$Sn$_{1-x}$Ge$_x$S$_6$ (\textit{x} = 0.1, 0.25, 1).

\vskip 0.25cm
\noindent
\textbf{\textit{Powder X-ray diffraction:}}  To confirm the structure and phase purity of the flux grown crystals, samples were analyzed with powder X-ray diffraction (PXRD).  Pieces of the crystals were ground by hand in a mortar and pestle to a fine powder, sifted through a 35 micron mesh sieve, and diffraction patterns were collected at room temperature on a Rigaku Miniflex-II instrument operating with Cu-K$\alpha$ radiation ($\lambda$ = 1.5406 Å) at 30 kV and 15 mA. The crystal structure was refined using the Rietveld method with GSAS-II software.\autocite{toby2013gsas}

\vskip 0.25cm
\noindent
\textbf{\textit{Magnetization:}} Temperature-dependent magnetization measurements were performed in Quantum Design Magnetic Property Measurement System (MPMS-classic and MPMS-3) SQUID magnetometers operating in the DC measurement mode and in a field 10 kOe. To avoid potential oxygen adsorption on the surface of the samples, several randomly oriented crystals (total mass
$\sim 80$ mg) of Ag$_8$SnS$_6$ were sealed in a partial argon atmosphere
between two fused silica rods of similar length in a fused silica tube.
The length of the rods and the tube were such that the only additional
contribution to the magnetic signal from the assembly comes
from the gap between the rods in which the samples are contained. This
contribution was subtracted using the size of the gap and previous
calibration. The powder samples were measured in a gel capsule which was mounted in a straw. To allow for evacuation, pinholes were poked in the straw and in the top of the capsule. For the powder samples, a separate background from an empty capsule was measured and subtracted. The measurements were conducted between 5--300 K with a heating/cooling rate of 0.5 K$\cdot$min$^{-1}$ between each temperature step. To study hysteresis and low temperature metastability, we also quenched (cooled as rapidly as possible, at $\approx$ 12 K$\cdot$min$^{-1}$) the sample to base temperature and measured on warming. In addition, time dependent measurements of the magnetization were performed at 40 K, 60 K and 80 K for up to 3,000 minutes (see Figure S4). These measurements were made after rapidly cooling ($\approx$ 12 K$\cdot$min$^{-1}$) the sample from 300 K to the target temperature (where the samples were cooled with the field applied).

\vskip 0.25cm
\noindent
\textbf{\textit{Thermal expansion:}} Measurements of the macroscopic length change of a crystal of Ag$_8$SnS$_6$ were performed using strain gauges. Similar to the design of Kabeya et al.,\autocite{kabeya2011thermal} the full setup used a Wheatstone configuration consisting of two strain gauges and two thin-film resistors. One strain gauge (type FLG-02-23, Tokyo Sokki Kenkyujo Co., Ltd. with R $\approx$ 120 $\Omega$) was fixed rigidly to the sample by using Devcon 5 minute epoxy (No. 14250). The second (identical) strain gauge was fixed to a sample of tungsten carbide, which has a comparatively small thermal expansion coefficient over a wide temperature range, to roughly balance out the intrinsic resistance changes of the strain gauges. In addition, two thin-film resistors with similar and almost temperature independent absolute resistances of $\approx$ 120 $\Omega$ were used to complete the Wheatstone bridge. The measured resistance changes were converted into relative length changes using the known gauge factor of the strain gauges (that was assumed to be temperature independent). The cryogenic environment was provided by a closed cycle cryostat (Janis SHI-950 with a base temperature of $\approx$ 3.5 K). Temperature control was provided by a LakeShore 336 temperature controller. Helium exchange gas was used to allow for fast cooldown times. The measurements were not aligned along a specific crystallographic direction.

\vskip 0.25cm
\noindent
\textbf{\textit{Structure determination:}} The crystal structure of Ag$_8$SnS$_6$ was determined at 295, 160 and 90 K by single crystal X-ray diffraction. Single crystal data were collected with a Bruker D8 VENTURE diffractometer (Photon CMOS detector, Mo-I$\mu$S microsource and Oxford Cryosystem 800 low temperature device). Data integration, absorption correction, and unit cell determination were performed with APEX 3 software. The starting atomic parameters were obtained by direct methods with the SHELXS-2017.\autocite{sheldrick2008short} Subsequently, the structures were refined using SHELXL-2017 (full-matrix least-squares on \textit{F$_o$}$^2$).

Single crystal X-ray diffraction data were first collected at 295 K. Afterwards, the sample was slowly cooled to 90 K at a rate 1 K$\cdot$min$^{-1}$, held at 90 K for 30 minutes, and a full data set was collected at 90 K. The next datasets were collected on heating at 160 K and 295 K. In both cases, the crystal was warmed at 6 K$\cdot$min$^{-1}$ and held at the target temperatures for 30 minutes prior to collection. Lastly, single crystal data was obtained on a crystal cooled as quickly as possible to 90 K. To accomplish this, the crystal was dismounted and rapidly (a few seconds) mounted back, directly into the nitrogen gas flow, allowing for fast cooling down to 90 K.

\vskip 0.25cm
\noindent
\textbf{\textit{Variable-temperature lattice parameters}}: The temperature dependent lattice constants were determined by tracking the room temperature (rt) orthorhombic $(8\,0\,4)_{\text{rt}}$, $(0\,4\,4)_{\text{rt}}$, and $(0\,0\,6)_{\text{rt}}$ Bragg peaks collected on single crystals. Measurements were performed on a four-circle diffractometer using Cu K$\alpha_1$ radiation generated from a rotating anode X-ray source using a germanium (1 1 1) monochromator. Scattered X-rays were measured using an FMB Oxford avalanche pixel diode (APD) single detector system. The sample was mounted on a copper pin using Crystal Bond. Sample was mounted in an ARS 4-800 K He-closed cycle refrigerator. The sample space is sealed with He exchange gas within the inner Be dome. Two further Be domes were mounted to act as a heat shield and the vacuum shroud.

Because during the growth process, the crystals are cooled through a known high temperature cubic to orthorhomic structural transition that occurs at 460 K\autocite{shen2020high}, the orthorhombic samples used in our experiments are highly twinned, i.e. cryptomorphic. The crystals form with facets perpendicular to the [1 1 1]$_{\text{PC}}$ (PC = pseudo-cubic) direction which in the rt orthorhombic structure split into the [2 0 1]$_{\text{rt}}$ and [0 1 1]$_{\text{rt}}$ directions. We aligned the crystal within the [0 1 1]$_{\text{rt}}$-[1 1 1]$_{\text{rt}}$ plane, which gives access to a large number of peaks from the available twin domains. Two peaks, the $(8\,0\,4)_{\text{rt}}$ and $(0\,4\,4)_{\text{rt}}$, were isolated from other peaks to give a measure of the \textit{a} and \textit{b} lattice parameters. The out of plane $(0\,0\,6)_{\text{rt}}$ peak was measured  to determine the \textit{c} lattice parameter. We first cooled from 300 K to 10 K at 0.2 K$\cdot$min$^{-1}$. The diffraction data used to determine the lattice parameters were collected by following the evolution of the above peaks on warming at 0.1 K$\cdot$min$^{-1}$, and the lattice parameters were calculated using Braggs' law.

\vskip 0.25cm
\noindent
\textbf{\textit{Computational Methods}}: Total energies and density of states (DOS) were calculated in density functional theory\autocite{hohenberg1964inhomogeneous,kohn1965self} (DFT) with PBEsol\autocite{perdew2008restoring} exchange-correlation functional, a plane-wave basis set and projected augmented wave\autocite{blochl1994projector} (PAW) method as implemented in VASP.\autocite{kresse1996efficient,kresse1996efficiency} A $\Gamma$-centered Monkhorst-Pack\autocite{monkhorst1976special} k-point mesh of (6$\times$6$\times$4) and (3$\times$6$\times$4) with a Gaussian smearing of 0.05 eV was used for the lt- and rt-structures with 30 and 60 atoms in the unit cells, respectively. A kinetic energy cutoff of 323.4 eV was used to fully relax the unit cell and atomic positions until the absolute force on each atom is below 0.02 eV/\AA. For bonding and anti-bonding analysis, the LOBSTER\autocite{maintz2013analytic} code was used to construct the crystal orbital Hamiltonian population (COHP) for nearest neighbor Ag-S pairs with the bond distance from 2.2 to 3.0 \AA.

\section{Results and Discussion}

\begin{figure}
    \centering
    \includegraphics[width=1\textwidth]{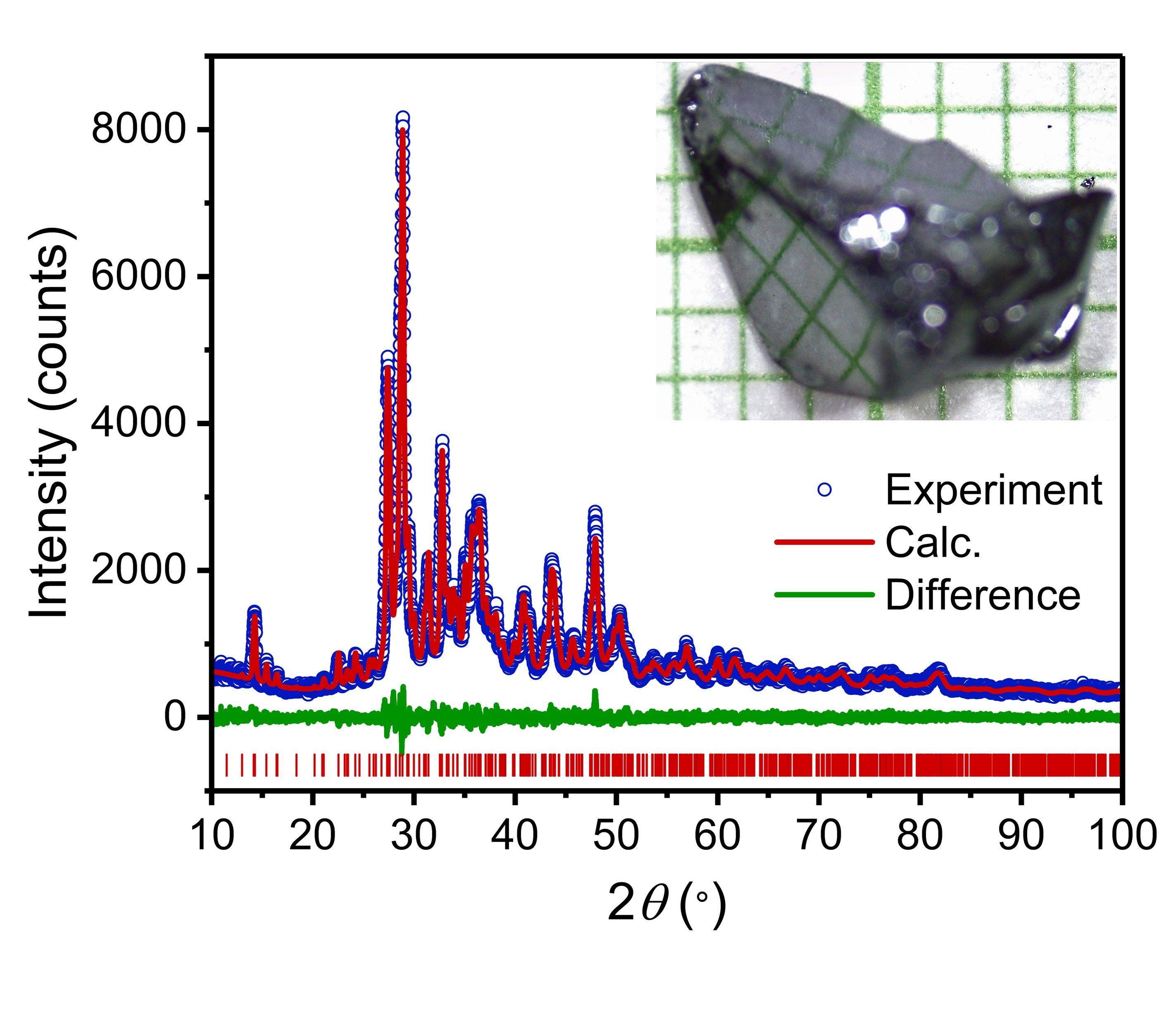}
    \caption{Powder X-ray diffraction pattern obtained on a powdered crystal of Ag$_8$SnS$_6$. The solid lines are the fitted results of a Rietveld refinement and are in good agreement with the room temperature \textit{Pna}2$_1$ structure of Ag$_8$SnS$_6$ with refinement statistics \textit{R}$_{\text{wp}}$ = 4.95 and GOF = 1.39. The inset is a picture of a flux grown crystal on a mm grid; the grid is reflected in two of the facets.}
    \label{Figure 1}
\end{figure}

The inset of Figure 1 displays an example of a single crystal grown from the Ag-Sn-S melt. To determine the identity and phase purity of the crystals, we analyzed samples with powder X-ray diffraction at room temperature. Figure 1 shows a representative powder pattern and refinement of the experimental data. The crystals adopt an orthorhombic structure with the \textit{Pna$2_1$} space group consistent with the room temperature (rt) phase of Ag$_8$SnS$_6$. The refined lattice constants are \textit{a} = 15.3155(9) \AA, \textit{b} = 7.5532(4) \AA, \textit{c} = 10.7000(6) \AA, in agreement with prior reports of Ag$_8$SnS$_6$.\autocite{li2000synthesis,mikolaichuk2010phase,shen2020high}  The experimental pattern has no observable peaks from secondary phases, supporting successful synthesis of phase-pure materials within the accuracy of our PXRD analysis. Close inspection revealed the crystals are heavily twinned, essentially cryptomorphic, forming with facets perpendicular to the pseudo-cubic [1 1 1]$_{\text{PC}}$ direction.  The twinning likely happens when cooling through the structural transition between the ht-cubic and rt orthorhombic phases that occurs at 460 K in Ag$_8$SnS$_6$.\autocite{shen2020high} 

Considering the known structural diversity among the broader argyrodite family and the limited low temperature data on Ag$_8$SnS$_6$, we explored the potential for phase transitions in Ag$_8$SnS$_6$ below room temperature. We first measured the magnetization of Ag$_8$SnS$_6$ between 5--300 K. The temperature dependent magnetic susceptibility (\textit{M/H}) is presented in Figure 2a, which shows \textit{M/H} is negative at all temperatures, indicating  diamagnetic behavior, and is consistent with our expectation of Ag$_8$SnS$_6$ being a valence precise semiconductor. At high temperatures, \textit{M/H} has a weak temperature dependence and increases slowly on cooling from -2.6$\times$10$^{-4}$ emu$\cdot$mol$^{-1}$ at 300 K to -2.55$\times$10$^{-4}$ emu$\cdot$mol$^{-1}$ at 120 K. The susceptibility data has an abrupt discontinuity at 120 K, falling from -2.55$\times$10$^{-4}$ emu$\cdot$mol$^{-1}$ to -2.62 $\times$10$^{-4}$ emu$\cdot$mol$^{-1}$. Given Ag$_8$SnS$_6$ is a moderate gap semiconductor with no local moment behavior expected or measured, the clear transition observed in \textit{M/H} suggests Ag$_8$SnS$_6$ undergoes a structural phase transition below $\approx$ 120 K.

\begin{figure}
    \centering
    \includegraphics[width=0.5\textwidth]{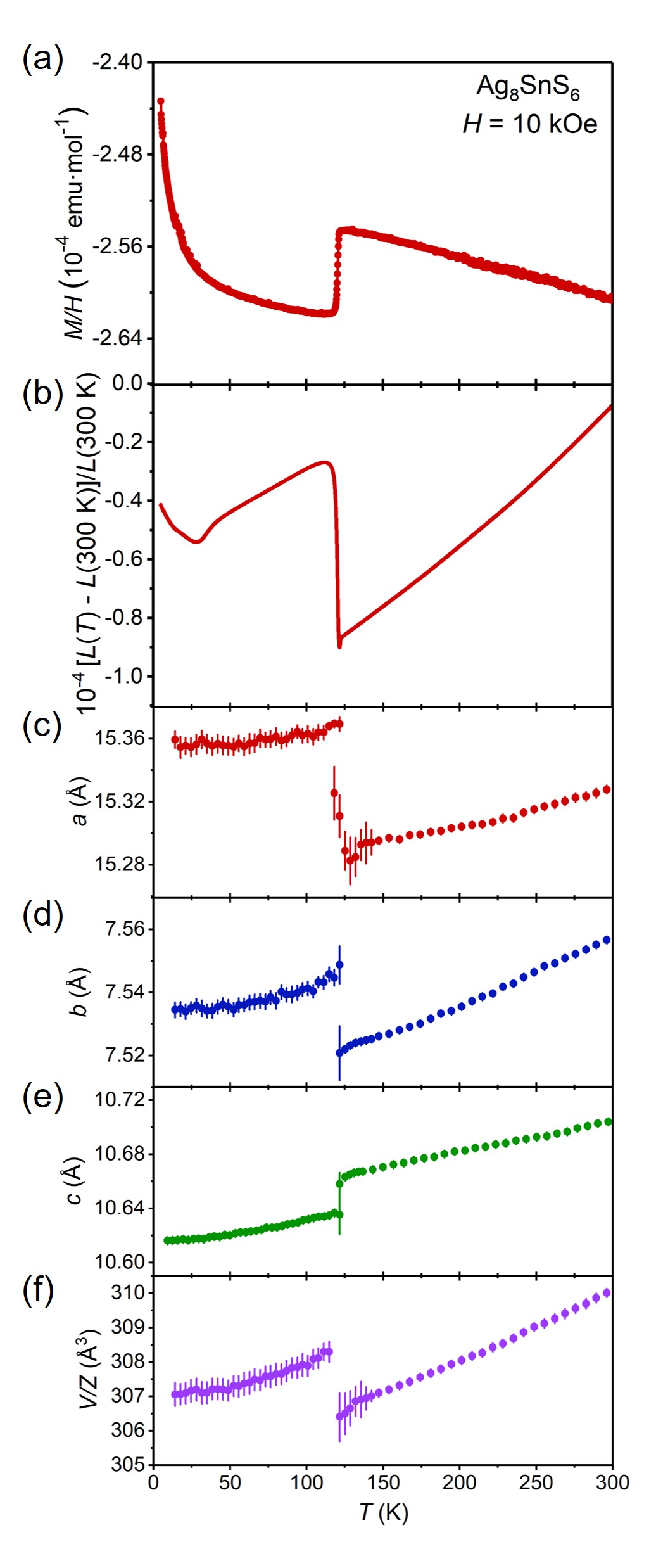}
    \caption{Temperature dependent physical properties of Ag$_8$SnS$_6$ single crystals. (a) Magnetic susceptibility \textit{(M/H)}, (b) Magnitude of the thermal expansion. (c–e) lattice parameters, and (f) unit cell volume. All datasets were measured on heating. The lattice parameters are shown using the basis of the rt phase (\textit{a}$_{rt}$ = 2\textit{a}$_{\text{lt}}$).}
    \label{Figure 2}
\end{figure}

To provide more detailed characterization of the phase transition, we tracked the thermal expansion of Ag$_8$SnS$_6$ between 5 and 300 K.  Figure 2b displays the measurement results, given as the magnitude of the temperature dependent length change normalized to the initial length at room temperature, [\textit{L}(\textit{T})-\textit{L}(300 K)]/\textit{L}(300 K) (referred to in the text as \textit{$\Delta$L/L})). As shown in Figure 2b, the values of \textit{$\Delta$L/L} show a rapid and large drop when warming above 120 K, implying the sample dimensions suddenly contract when heating above this temperature. The sudden, jump-like, length change suggests a first-order phase transition, and is characteristic of a structural transformation. The transition temperature closely matches what is observed in the magnetization data, and together the two measurements strongly suggest Ag$_8$SnS$_6$ undergoes a first-order structural transition at 120 K. The feature near 30 K in Figure 2b is an artifact caused by the slightly different temperature dependencies of the resistances of the strain gauges (used to determine \textit{$\Delta$L/L}) at low temperatures.

Having found evidence for a previously unknown structural phase transition in Ag$_8$SnS$_6$, we performed single crystal X-ray diffraction at 295 K, 160 K and 90 K to determine the crystal structure of the low temperature (lt) phase. Details regarding the structural refinements are given in the Supporting Information in Table S1, and the crystallographic information regarding the atomic positions, thermal displacement parameters, and individual Ag-S and Sn-S bond lengths are provided in Tables S2-S11. Figure 3 compares the room and low temperature structures of Ag$_8$SnS$_6$ as determined by our single crystal diffraction analysis conducted at 295 and 90 K. 

The first dataset collected at 295 K agrees with the expected rt orthorhombic unit cell with space group \textit{Pna}2$_1$ ($\#$33) and \textit{a} = 15.2993(9) \AA, \textit{b} = 7.5479(4) \AA, \textit{c} = 10.7045(6) \AA, and phase volume \textit{V/Z} = 309.03(3) \AA$^3$, where \textit{Z} = 4 is the number of formula units per unit cell in the rt structure. These results are consistent with earlier literature.\autocite{wang1978new,shen2020high} The rt structure is complex, as illustrated by Figures 3a and 3b. All the atoms occupy the Wyckoff position 4a and are labeled respectively as Ag1-Ag8, Sn, and S1-S6. The local environments of the Ag atoms are diverse, with Ag found in distorted tetrahedra (Ag2 and Ag6) as well as trigonal planar (Ag1, Ag3, Ag4, and Ag7) and linear (Ag5) coordination by S. As given in Table S10 of the Supporting Information, the Ag coordination environments are all distorted from the ideal geometries with a wide span of Ag-S bond lengths ranging from 2.4 to 2.69 \AA. Each Sn atom is likewise coordinated by a distorted tetrahedron of S atoms with bond lengths of approximately 2.37 \AA. The Ag-S and Sn-S tetrahedra form a complicated three dimensional corner sharing network, with some tetrahedral vertices also linked by the linear and trigonal planar coordinated Ag atoms as shown in Figure 3a and 3b.  

\begin{figure}
    \centering
    \includegraphics[width=1\textwidth]{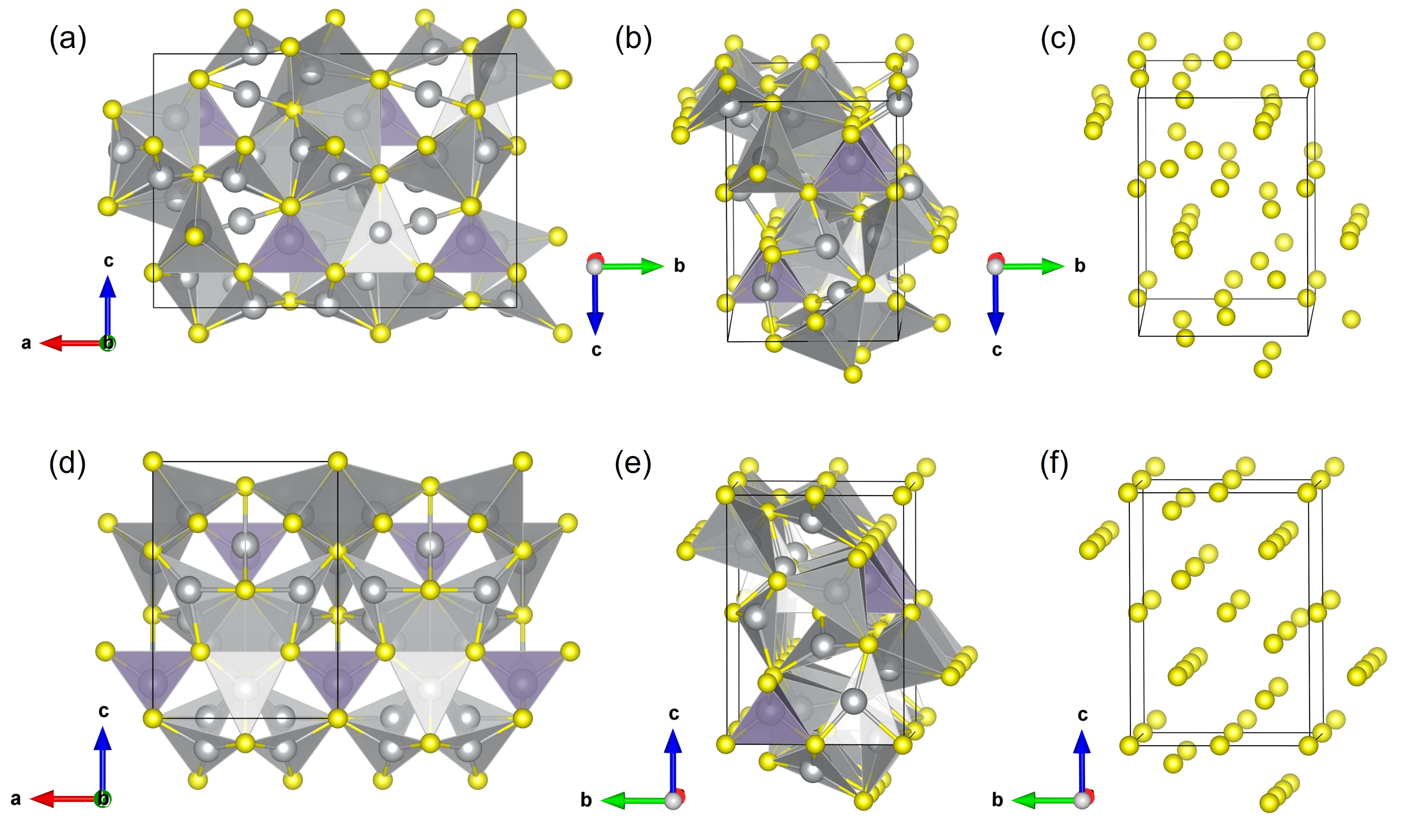}
    \caption{Crystal structure of Ag$_8$SnS$_6$ determined at 295 K and 90 K. The 295 K (rt) unit cell viewed along the (a) \textit{b}-axis and (b) slightly titled off the \textit{a}-axis. (c) The same perspective of (b) showing only the S atoms. (d-f) show the same for the 90 K (lt) structure with the unit cell doubled along the \textit{a}-axis for comparison with the rt phase. The perspectives in (c) and (f) emphasize the straightening of the S columns along the \textit{a}-axis in the lt polymorph. The color code is as follows: Ag: grey, Sn: purple, S: yellow.}
    \label{Figure 3}
\end{figure}

Analysis of the 90 K diffraction data confirms the structural phase transition. We find that at low temperatures, Ag$_8$SnS$_6$ adopts a different orthorhombic structure with the space group \textit{Pmn}2$_1$ ($\#$ 31) and cell parameters \textit{a} = 7.6629(5) \AA, \textit{b} = 7.5396(5) \AA, \textit{c} = 10.6300(5) \AA, \textit{V/Z} = 307.08(4) \AA$^3$ (Z = 2 in the lt structure). The lt polymorph of Ag$_8$SnS$_6$ is isostructural to the room temperature forms of the related Se-based argyrodite compounds Ag$_8$SnSe$_6$ and Ag$_8$GeSe$_6$.\autocite{gulay2002crystal,semkiv2017ag8snse6,carre1980structure} Several perspectives on the low temperature variant of Ag$_8$SnS$_6$ are illustrated in Figures 3d-f. We show the lt structure with the unit cell doubled along \textit{a} for ease of comparison with the rt structure. In the lt polymorph, Ag atoms occupy three 4b and two 2a wyckoff sites, Sn – one 2a wyckoff site, and S – one 4b wyckoff site. As in the rt phase, there is one Sn atom occupying a distorted tetrahedron of S atoms with bond lengths of approximately 2.38 \AA. The lt unit cell contains five unique Ag atoms, three of which (Ag1, Ag2, and Ag5) are found in acentric trigonal planar coordination by sulfur and two (Ag3 and Ag4) that are tetrahedrally coordinated. The Ag–S bond lengths again cover a wide range from 2.44 to 2.79 \AA\ (see Table S10 in the Supporting Information). Unlike the rt phase, the lt structure does not contain Ag atoms in linear coordination.  

Figures 3b and 3e illustrate how the room and low temperature structures have similar connectivity of atoms, manifested most clearly in the chains/columns of S atoms running down the \textit{a} axis. We emphasize this in Figures 3c and 3f by showing only the positions of the S atoms in each polymorph. In the rt phase, the S columns are slightly nonlinear, but in the lt structure the chains straighten such that the S atoms overlay when viewed down the \textit{a}-axis. In this light, the rt structure can be viewed as a distortion of the lt phase, in which the new periodicity resulting from the bending of the S chains doubles the unit cell along the \textit{a} direction. Notably, \textit{Pna}2$_1$ ($\#$33) is the maximal non-isomorphic “klassengleich” (class-equivalent) subgroup with index of 2 (k2) of the space group \textit{Pmn}2$_1$ ($\#$31).\autocite{hahn1983international} Therefore, upon transitioning from lt to rt phase on warming, the primitive-orthorhombic cell is retained, but the volume is doubled due to the doubling of the \textit{a}-lattice parameter. After the transition, each 2a site in the lt structure transforms into a single 4a site in the rt structure, while the 4b site in lt structure becomes two 4a sites in rt structure. This relationship is somewhat unusual, differing from the more common reduction of symmetry on cooling. Comparing the three polymorphs of Ag$_8$SnS$_6$, the crystal structure becomes increasingly distorted as the temperature rises. Starting from the lt \textit{Pmn}2$_1$ arrangement, the linear S chains become bent on transitioning into the rt \textit{Pna}2$_1$ structure, and this is accompanied by an increase in the diversity of local coordination environments adopted by the Ag atoms. Finally, the Ag sublattice enters a highly disordered state in the ht cubic phase with high ionic conductivity, where the Ag atoms are distributed over three partially occupied sites.

After determining the lt crystal structure, we next measured the temperature dependence of the lattice parameters by following the room temperature (rt) orthorhombic (8 0 4)$_{\text{rt}}$, (0 4 4)$_{\text{rt}}$, and (0 0 6)$_{\text{rt}}$ diffraction peaks collected on single crystals. Figure 4 shows the evolution of the Bragg peaks on warming in the vicinity of the transition, between $\approx$ 100-140 K. Each peak shows sudden changes as the temperature rises above $\approx$ 120 K, indicative of the phase change. The (0 0 6)$_{\text{rt}}$ peak shifts to lower 2$\theta$ by $\approx$ 0.1°. The (8 0 4)$_{\text{rt}}$ reflection moves to higher 2$\theta$ and is significantly reduced in intensity. The (0 4 4)$_{\text{rt}}$ peak moves to higher 2$\theta$ and has strongly enhanced intensity. Close inspection of the data collected at $\approx$ 120 K shows reflections corresponding to each structure, implying coexistence of both phases across the transition region, which is characteristic of a first-order transformation (the peaks corresponding to each phase are marked by arrows in Figure 4). 

\begin{figure}[!h]
    \centering
    \includegraphics[width=1\textwidth]{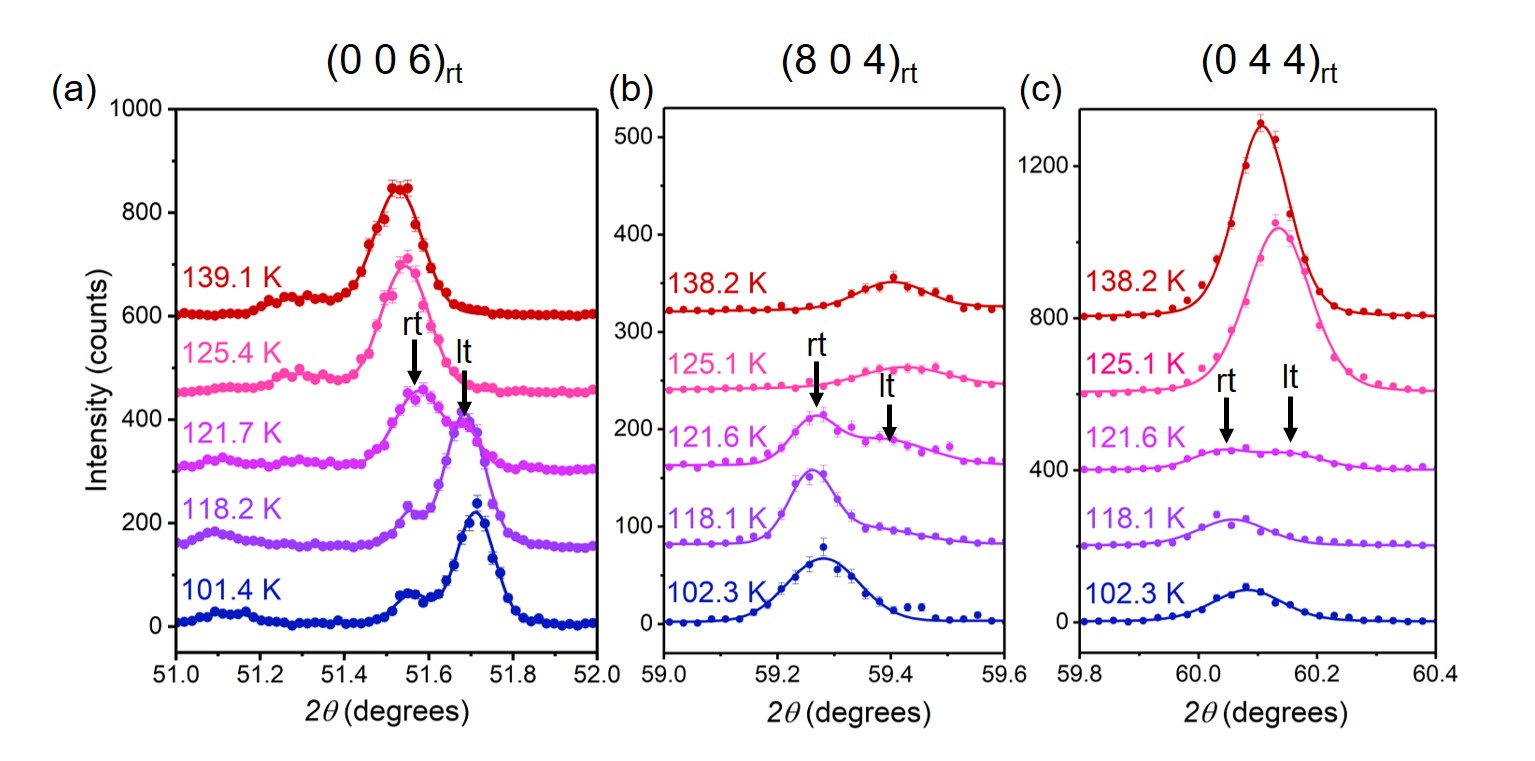}
    \caption{Temperature dependence of the (a) (0 0 6)$_{\text{rt}}$, (b) (8 0 4)$_{\text{rt}}$, (c) and (0 4 4)$_{\text{rt}}$ Bragg peaks of a Ag$_8$SnS$_6$ single crystal measured upon warming between $\approx$100--140 K on 4-circle diffractometer.}
    \label{Figure 4}
\end{figure}

Figures 2c-e display the temperature dependent lattice parameters extracted from the data. The lattice parameters all show abrupt discontinuities at 120 K, consistent with the magnetization and thermal expansion measurements. On warming from the base temperature, the \textit{a}-, \textit{b}-, and \textit{c}-lattice parameters all increase weakly up to approximately 120 K, after which the \textit{a}- and \textit{b}-lattice parameters rapidly contract and the \textit{c}-lattice parameter expands. The unit cell dimensions and phase volume all increase gradually with further heating above 120 K. The values at room temperature and 90 K agree with those determined by the single crystal structural refinements discussed above. We emphasize here that below the structural transition, \textit{a}$_{\text{rt}}$ $\approx$ 2\textit{a}$_{\text{lt}}$ and \textit{Z}$_{rt}$/\textit{Z}$_{lt}$ = 2 in the lt phase. The contraction along \textit{a} shown in Figure 2c implies a shrinking of the d-spacing along the \textit{a}-axis on heating above 120 K. The net result is a decrease in \textit{V/Z} from 308 Å$^3$ to 306 Å$^3$ when warming across the transition at 120 K (see Figure 2f). We also measured variable temperature powder XRD; however, the structural transition becomes either completely arrested or broadened in the powdered samples. The variable temperature powder XRD data are presented and discussed in the Supporting Information (Figure S1). 

Given that the transition is structural and appears to be first-order in nature, we also collected magnetization and thermal expansion data upon heating and cooling. The magnetic data is shown in Figure 5a, demonstrating large hysteresis in \textit{M/H} between heating and cooling at 0.5 K$\cdot$min$^{-1}$.  Comparable hysteresis is also detected in the thermal expansion data, as displayed in Figure 5b. Since the extent of the hysterisis can depend on the rate of temperature change, we quenched our samples (i.e. cooled as rapidly as possible to 5 K) and then measured upon slowly warming. For both magnetization and thermal expansion measurements, the trend from 5-60 K (after quenching) appears to match a rough extrapolation of the 130-300 K data to lower temperatures. As the sample is heated above 60 K, both magnetization and thermal expansion datasets show a clear drop followed by a rapid increase on warming at 120 K. These results imply that fast cooling traps the crystal in the rt phase. The metastable state is maintained upon slow warming until about 60-70 K, where both the \textit{M/H} and \textit{$\Delta$L/L} data return to the slow heating/cooling manifold. 

\begin{figure}
    \centering
    \includegraphics[width=0.7\textwidth]{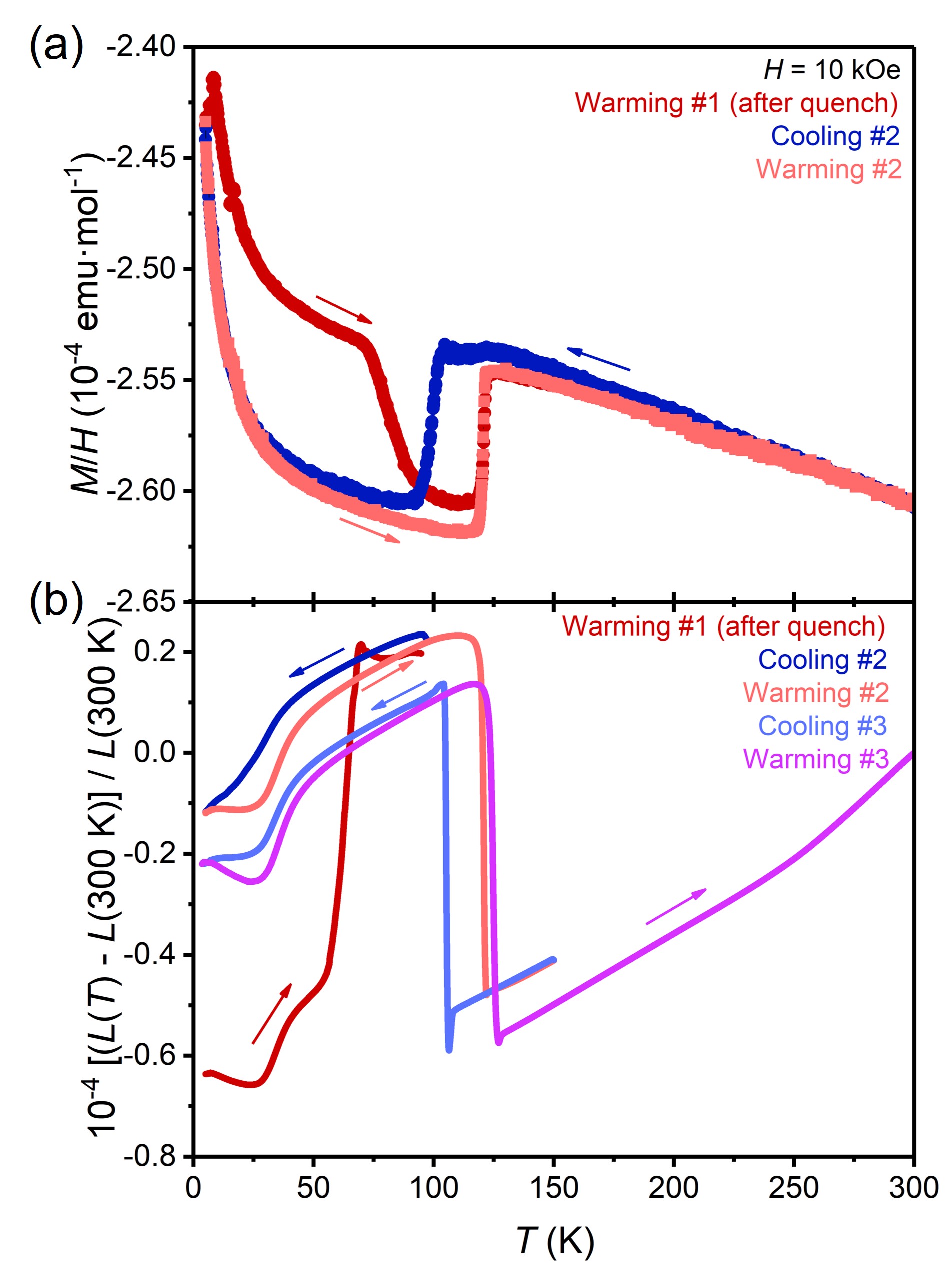}
    \caption{Thermal hysteresis of the structural transition in Ag$_8$SnS$_6$. (a) Temperature dependent magnetic susceptibility and (b) thermal expansion. See text for details about quenching.}
    \label{Figure 5}
\end{figure}

We furthermore measured the thermal expansion with different heating and cooling rates. The results are presented in Figure S3 in the Supporting Information, and indicate cooling at a rate faster than 4-5 K/min is sufficient to arrest the crystallographic transformation. On warming, a clear transition to the lt polymorph begins at $\approx$ 60 K, and this transformation is relatively insensitive to the heating rate. Given that the metastable state appears to be preserved below 60 K, we lastly tracked the magnetic susceptibility over time at \textit{T} = 40 K, 60 K, and 80 K to determine how quickly the thermodymically stable lt variant is recovered after quenching to each temperature. Figure S4 in the supporting information shows the results. We find that the transition into the lt polymorph begins almost immediately at 80 K, takes 100's of minutes to become noticeable at 60 K, and remains in the metastable rt phase for the duration of our 1000 minute measurement at 40 K. 

The lt-rt phase transition can be considered as a reconstructive phase transition,\autocite{buerger1961polymorphism,Buerger} e.g. chemical bonds are broken and re-formed. Such transitions involve considerable atomic motions and are always first-order with hysteresis. The hysteresis can be understood as a case of kinetic arrest,\autocite{chattopadhyay2005kinetic} caused by the competition between the Gibbs free energy (the transformation driving force), the strain energy (transformation opposing force), and the thermal energy that allows for atomic rearrangement. Our thermal expansion and magnetization data suggest the thermal energy below at least 40 K is insufficient to drive the structural rearrangement to the lt polymorph, allowing for the sample to be arrested in the rt structure by rapid cooling. On warming the quenched sample, the structure transforms to the thermodynamically favored lt structure when the thermal energy overcomes the activation barrier. These results suggest that Ag$_8$SnS$_6$ may serve as a model system for the study of time-temperature-phase relations associated with a quenchable, structural phase transition. The advantages of Ag$_8$SnS$_6$ are (i) it is a stoichiometric phase that can be grown in high purity form and (ii) the transition from one phase to another can be monitored by a contact-free measurement such as magnetization.

Although we were not able to perform single crystal X-ray diffraction studies below 90 K, i.e we could not reach the quenched, metastable rt structure that exists for long periods of time below 40 K, we could still directly compare the two orthorhombic structures (rt and lt versions) at the same temperature by taking advantage of the hysteresis of the phase transition (i.e. in the 90--125 K region in Figure 5). Given that the width the hysteresis depends on cooling rate (Figure S3), we cooled a sample as rapidly as possible to 90 K and collected single crystal diffraction data (see the Experimental Section for more details). After fast cooling, the diffraction data revealed an orthorhombic \textit{Pna}2$_1$ unit cell (\textit{a} = 15.2724(9) \AA, \textit{b} = 7.5197(5) \AA, \textit{c} = 10.6569(7) \AA, \textit{V/Z} = 305.98(3) \AA$^3$), consistent with rt Ag$_8$SnS$_6$. The quickly cooled, 90 K, results are in qualitative agreement with the behavior of the lattice parameters shown in Figure 2c-f. Table S10 in the Supporting Information compares Sn-S and Ag-S bond lengths determined by the single crystal data at each temperature, and we find that at 90 K, most bonds are slightly longer in the lt variant than in the 90 K rt \textit{Pna}2$_1$ arrangement, consistent with the larger \textit{V/Z} of the lt structure.

Like rt Ag$_8$SnS$_6$, the related argyrodite compounds Ag$_8$GeS$_6$ and Ag$_8$SiS$_6$ both adopt the \textit{Pna}2$_1$ structure,\autocite{krebs1977kenntnis,eulenberger1977kristallstruktur} whereas Ag$_8$SnSe$_6$ and Ag$_8$GeSe$_6$ take the \textit{Pnm2$_1$} form.\autocite{gulay2002crystal,semkiv2017ag8snse6,carre1980structure}  From this information, we observe a trend relating the unit cell dimensions (normalized to \textit{Z}) to the preferred structure of each material. Owing to the smaller size of Ge and Si compared to Sn, Ag$_8$GeS$_6$ and Ag$_8$SiS$_6$ have smaller respective phase volumes of 299 and 296 \AA$^3$ compared to 309 \AA$^3$ for Ag$_8$SnS$_6$. Similarly, Se is larger than S, and Ag$_8$GeSe$_6$ and Ag$_8$SnSe$_6$ both have larger \textit{V/Z} of 328 and 342 \AA$^3$ respectively. The preferred structures indicate that increasing \textit{V/Z} favors the \textit{Pmn}2$_1$ arrangement. This reasoning is consistent with the intermediate \textit{V/Z} of Ag$_8$SnS$_6$ and the observation of both polymorphs in this compound.

To support this argument based on crystallography, we used DFT to compare the calculated formation energies of Ag$_8$SnS$_6$ and Ag$_8$GeS$_6$ in both \textit{Pna}2$_1$ and \textit{Pmn}2$_1$ structures at zero temperature. With the PBEsol exchange-correlation functional, the fully-relaxed lattice parameters for Ag$_8$SnS$_6$ are \textit{a} = 15.03 \AA, \textit{b} = 7.40 \AA, and \textit{c} = 10.58 \AA\  for the rt structure, and \textit{a} = 7.63 \AA, \textit{b} = 7.42 \AA, and \textit{c} = 10.49 \AA\ for the lt structure. Both relaxed lattice parameters are within 2$\%$ (smaller) than the experimental data, which is typically acceptable for PBEsol. The corresponding total energy difference ($\Delta$\textit{E}) favors the lt structure by 0.14 eV$\cdot$f.u.$^{-1}$ (or 9.5 meV$\cdot$atom$^{-1}$), agreeing with the experimental observation of the \textit{Pnm}2$_1$ variant being the ground state structure. For Ag$_8$GeS$_6$, the calculations indicate the \textit{Pnm}2$_1$ structure is also more stable, but by only 2.9 meV$\cdot$atom$^{-1}$. We note that the calculated $\Delta$\textit{E} is only weakly affected by the inclusion of spin-orbit coupling (SOC), changing to 9.7 meV$\cdot$atom$^{-1}$ for Ag$_8$SnS$_6$. Thus, the following results are all from DFT calculations without SOC. 

The calculations are qualitatively consistent with our suggestion that the \textit{Pmn}2$_1$ structure is favored to a greater degree for larger phase volume. This implies that applying pressure to Ag$_8$SnS$_6$, or contracting the lattice, should suppress the transition temperature. We tested this by applying chemical pressure through isovalent alloying with the smaller Ge in place of Sn. We prepared single crystals of Ag$_8$Sn$_{1-x}$Ge$_x$S$_6$ (nominal \textit{x} = 0.10, 0.25) and pure Ag$_8$GeS$_6$ following the same growth procedure used for the Ag$_8$SnS$_6$. Powder X-ray diffraction analysis indicate successful Ge incorporation (Figures S4 and S5 in the Suporting Information), with a monotonic decrease in the lattice parameters and unit cell volume moving from pure Ag$_8$SnS$_6$ to Ag$_8$GeS$_6$. Magnetization data measured from 5 to 300 K is given in the Supporting Information in Figure S7, and we find no evidence for a phase transition in any of the Ge containing samples. Although we cannot definitively rule out a disorder broadened transition, the magnetization data supports the arguments presented above, and indicates the lattice contraction caused by even relatively small ($\approx$10$\%$) addition of Ge is sufficient to eliminate the transition. Suppression of the transition after lattice contraction is likewise consistent with our X-ray analysis, which showed an increase in \textit{V/Z} when cooling through the transition. In principle, our arguments could be further tested by expanding the lattice with "negative pressure" or alloying with larger atoms, which we anticipate would raise the transition temperature of Ag$_8$SnS$_6$. Likewise, we predict that applying pressure to Ag$_8$GeSe$_6$ may stabilize a currently unreported \textit{Pna}2$_1$ polymorph.

Our calculations and experimental results indicate the preferred structure of the agryrodite members is closely related to the normalized unit cell volume. This observation is likely grounded in the different strength of the chemical bonding in each material, which will naturally be reflected in the bond lengths. The structures adopted by the various argyrodite materials, along with the experiments and calculations presented here, suggest larger volume (longer bonds) favors the \textit{Pmn}2$_1$ variant. Because the local tetrahedral coordination environment around Sn remains essentially unchanged in each phase, whereas the changes to the Ag-S sublattice are more significant, we anticipate the Ag-S interactions are most important in determining the preferred structure. 

To provide more insight into the role of chemical bonding in determining the relative stability for the two Ag$_8$SnS$_6$ structures, we also used DFT to calculate the density of states (DOS) and crystal orbital Hamilton population (COHP) for each polymorph. Figure 6a shows the calculated total DOS for each structure of Ag$_8$SnS$_6$, and Figure 6b displays the atomic orbital projected density of states (PDOS) for the lt structure to show the positions of the relevant bands. The overall DOS for each structure are very similar. The states from -14 to -12 eV are mostly from the S 3s orbitals and those from -8 to -7 eV are primarily from Sn 5s orbitals, respectively. The Sn 5p bands (plotted together with Sn 5s) spread out in a broad energy range above -6 eV and have low magnitude due to the small number of Sn atoms. The Ag 4d bands range from -6 and 0 eV, with the main peak centered around -3.5 eV. The Ag 5s bands spread from -6 eV to the conduction band, with greatest contribution from -6 to -5.5 eV, where they overlap with the lower edge of the Ag 4d bands. The S 3p bands range from -8 eV to the conduction band with two regions of greatest contribution, one from -6 to -4.5 eV and the other from -2 to 0 eV. 

\begin{figure}
    \centering
    \includegraphics[width=1\textwidth]{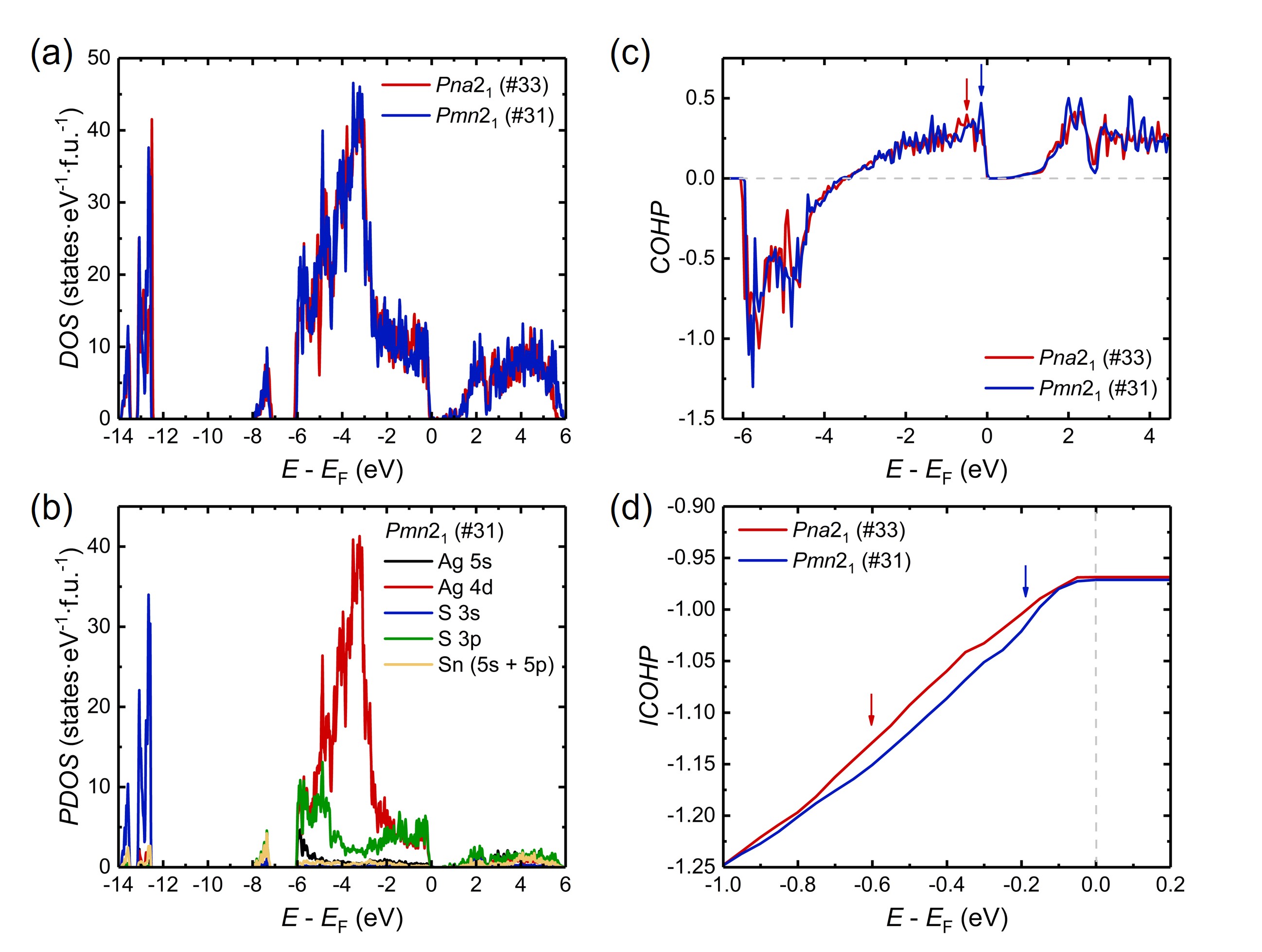}
    \caption{(a) DFT calculated density of states for Ag$_8$SnS$_6$ in the rt (\textit{Pna}2$_1$) and lt (\textit{Pmn}2$_1$) structure. (b) Orbital projected density of states for Ag$_8$SnS$_6$ in the lt structure. (c) Crystal orbital Hamiltonian population (COHP) plots for nearest neighbor Ag-S interactions in Ag$_8$SnS$_6$ for rt  and lt structures. (d) Integrated COHP for each structure. The arrows in (c) and (d) mark the positions of relevant antibonding COHP peaks that shift to higher energy in the lt structure.}
    \label{Figure 6}
\end{figure}

The COHP for nearest-neighbor Ag-S pairs is presented in Figure 6c. COHP decomposes the density of states into nearest neighbor pair interactions and gives information on bonding and antibonding contributions to the band structure energy,\autocite{dronskowski1993crystal} respectively denoted by negative (stabilizing) and positive (destabilizing) COHP values. The COHP plot clearly shows the Ag-S band hybridization from -6 to -3.4 eV gives bonding states and that the states above -3.4 eV are anti-bonding. The relative stability of the two structures depends on the details of how the states in these bonding and anti-bonding regions distribute differently, which reflects the rearrangements of Ag-S local coordination and the different bond lengths in each structure. In the bonding region, the primary COHP peaks reach more negative values in the lt structure. Likewise, in the antibonding region, the most notable change is the shift of the COHP peak from -0.6 to -0.2 eV as indicated by the two vertical arrows in Figure 6c. The effects of these changes on the overall stability are captured by integrating the COHP (ICOHP) to the Fermi level, which gives the contribution from the nearest neighbor Ag-S pairs to the total energy. As shown in Figure 6d and zoomed just below the Fermi energy \textit{E}$_F$, after integrating up to \textit{E}$_F$, we ultimately find the lt arrangement to be more stable than the rt-structure by 2 meV/pair, in qualitative agreement with our overall DFT energy calculations discussed above. The calculations therefore indicate differences in the Ag-S bonding significantly impact the energetically favored structure.  

\section{Summary and Conclusions}

We grew single crystals of the semiconducting mineral Ag$_8$SnS$_6$ from a Ag-Sn-S melt. Temperature dependent magnetization, thermal expansion, and X-ray diffraction measurements demonstrate Ag$_8$SnS$_6$ crystals undergo a structural transition from the room temperature (\textit{Pna}2$_1$) form to a different orthorhombic (\textit{Pmn}2$_1$) structure below 120 K (on warming). The low temperature polymorph is isostructural to the related argyrodites Ag$_8$SnSe$_6$ and Ag$_8$GeSe$_6$.  We find the transition temperature to be strongly dependent on the cooling rate, and samples rapidly quenched at speeds faster than $\approx$ 4-5 K$\cdot$min$^{-1}$ can be trapped in a metastable form of the room temperature phase for temperature up to at least 40 K. As such, Ag$_8$SnS$_6$ may serve as a model system for the study of time-temperature-phase relations associated with a quenchable, structural phase transition.  Based on the structures preferred by each family of argyrodite materials Ag$_8$\textit{TQ}$_6$ (\textit{T} = Si, Ge, Sn; \textit{Q} = S, Se), we suggest lower volumes (per formula unit) increasingly favor the \textit{Pna}2$_1$ variant. We support this picture by applying chemical pressure to Ag$_8$SnS$_6$ by alloying with Ge to form Ag$_8$Sn$_{1-x}$Ge$_x$S$_6$, and find that a small lattice contraction from 10 $\%$ incorporation of Ge is sufficient to avoid the low temperature phase change in Ag$_8$Sn$_{1-x}$Ge$_x$S$_6$. We lastly use density function theory calculations to suggest that differences in Ag-S bonding determines the energetically preferred structure.

\vskip 0.25cm
\noindent
\textbf{\textit{Acknowledgements}}

Work at the Ames Laboratory (T.J.S., J.M.W., A.K., E.G., L.W., R.A.R., Y.M., V.K.P., S.L.B., P.C.C.) was supported by the U.S. Department of Energy, Office of Science, Basic Energy Sciences, Materials Sciences and Engineering Division. The Ames Laboratory is operated for the U.S. Department of Energy by Iowa State
University under Contract No. DE-AC02-07CH11358. T.J.S. was supported by the Center for Advancement of Topological Semimetals (CATS), an Energy Frontier Research Center funded by the U.S. Department of Energy Office of Science, Office of Basic Energy Sciences, through the Ames Laboratory under its Contract No. DE-AC02-07CH11358 with Iowa State University. T.J.S and E.G. also acknowledge funding from the Gordon and Betty Moore Foundation’s EPiQS Initiative through Grant No. GBMF4411. J.V.Z. acknowledges financial support from the National Science Foundation (DMR 1944551). Temperature dependent powder diffraction experiments (Y.M., V.K.P.) were supported by the Division of Materials Science and Engineering of the Office of Basic Energy Sciences.  Earlier attempts to grow and study single crystal Ag$_8$SnS$_6$ samples were made by X. Lin,  W. Meier and G. Drachuck. P.C.C. acknowledges D. Argyriou for geologic reasons. The authors thank Tom Lagrasso and Matt Kramer for useful discussions.

\vskip 0.25cm
\noindent
\textbf{\textit{Conflicts of Interest}}

The authors have no conflicts of interest to declare.

\end{doublespace}

\newpage

\printbibliography

\end{document}

\begin{center}
    \huge{Supporting Information} \\[10pt]
    \vskip 0.25cm
    \huge{A Low Temperature Structural Transition in Canfieldite, Ag$_8$SnS$_6$, Single Crystals}
    \vskip 0.25cm
    \large{Tyler J. Slade$^{1,2}$, Volodymyr Gvozdetskyi$^3$, John M. Wilde$^{1,2}$, Andreas Kreyssig$^{1,2}$, Elena Gati$^{1,2}$, Lin-Lin Wang$^{1,2}$, Yaroslav Mudryk$^{1}$, Raquel A. Ribeiro$^{1,2}$, Vitalij K. Pecharsky$^{1,4}$, Julia V. Zaikina$^3$, Sergey L. Bud’ko$^{1,2}$, Paul C. Canfield$^{1,2}$} \\[3pt]
\end{center}

\textit{$^{1}$Ames Laboratory, US DOE, Iowa State University, Ames, Iowa 50011, USA} \\
\textit{$^{2}$Department of Physics and Astronomy, Iowa State University, Ames, Iowa 50011, USA}
\\
\textit{$^{3}$Department of Chemistry, Iowa State University, Ames, Iowa 50011, USA}
\\
\textit{$^{4}$Department of Materials Science and Engineering, Iowa State University, Ames, IA 50011, USA}

\clearpage
\newpage

\begin{doublespace}

\vskip 0.25cm
\noindent
\textbf{\textit{Absence of the low temperature structural transition in powdered Ag$_8$SnS$_6$}} After identifying in Ag$_8$SnS$_6$ a phase change from \textit{Pna}2$_1$ to \textit{Pnm}2$_1$ structures, we attempted to use variable temperature powder X-ray diffraction (PXRD) to analyze the structural changes while cooling and heating through the transition and give an accurate measurement of the temperature dependent lattice parameters. The powder diffraction results are shown below in Figure S1. Data collected on warming from 10 K are shown in Figure S1a and patterns obtained on cooling in Figure S1b. Both figures also show the simulated diffraction patterns for Ag$_8$SnS$_6$ in the rt \textit{Pna}2$_1$ and lt \textit{Pnm2$_1$} structures. At each temperature, the experimental data is clearly in better agreement with the rt \textit{Pna}2$_1$ structure, with the largest discrepancies at low angles, where two peaks anticipated for the lt structure are absent in the experimental data (see the light blue arrows in Figure S1). Rietveld refinements of the same data produced satisfactory refinements (\textit{R}$_{\text{wp}}$ $<$ 10) of every pattern when using the rt \textit{Pna}2$_1$ structural model, while attempts to refine the experimental data using the \textit{Pmn}2$_1$ structure gave poor fits (results not shown). Therefore, the variable temperature PXRD results give no evidence of a transformation to the \textit{Pmn}2$_1$ polymorph.

To help understand the variable temperature PXRD data and place it in better experimental context with our single crystal diffraction results, we likewise measured magnetization on a powdered sample and present the data in Figure S2. Like the powder diffraction results, the phase transition seen in the crystals is not observed in \textit{M}(\textit{T}) for the powdered sample. The \textit{M}(\textit{T}) data does, however, show a broad feature between 50-70 K that is only observed on heating. The peak is reproducible in multiple measurements on different powdered samples. Notably, a similar feature was observed on warming and at the same temperatures in some of our measurements on single crystals (data not shown). In the single crystal data, we could eliminate the peak by carefully ensuring no oxygen was present during the measurements (see the description of the magnetic measurements in the experimental section). Therefore, we believe this peak is most likely from trace oxygen that proved more difficult to eliminate in the powder. Overall, both magnetization and PXRD data indicate the structural transition does not occur in powdered samples, possibly because of the strain introduced during grinding. The absence of a transition in powdered samples is not further discussed in the present work.

\clearpage
\newpage

\begin{figure}[!h]
    \centering
    \includegraphics[width=0.8\textwidth]{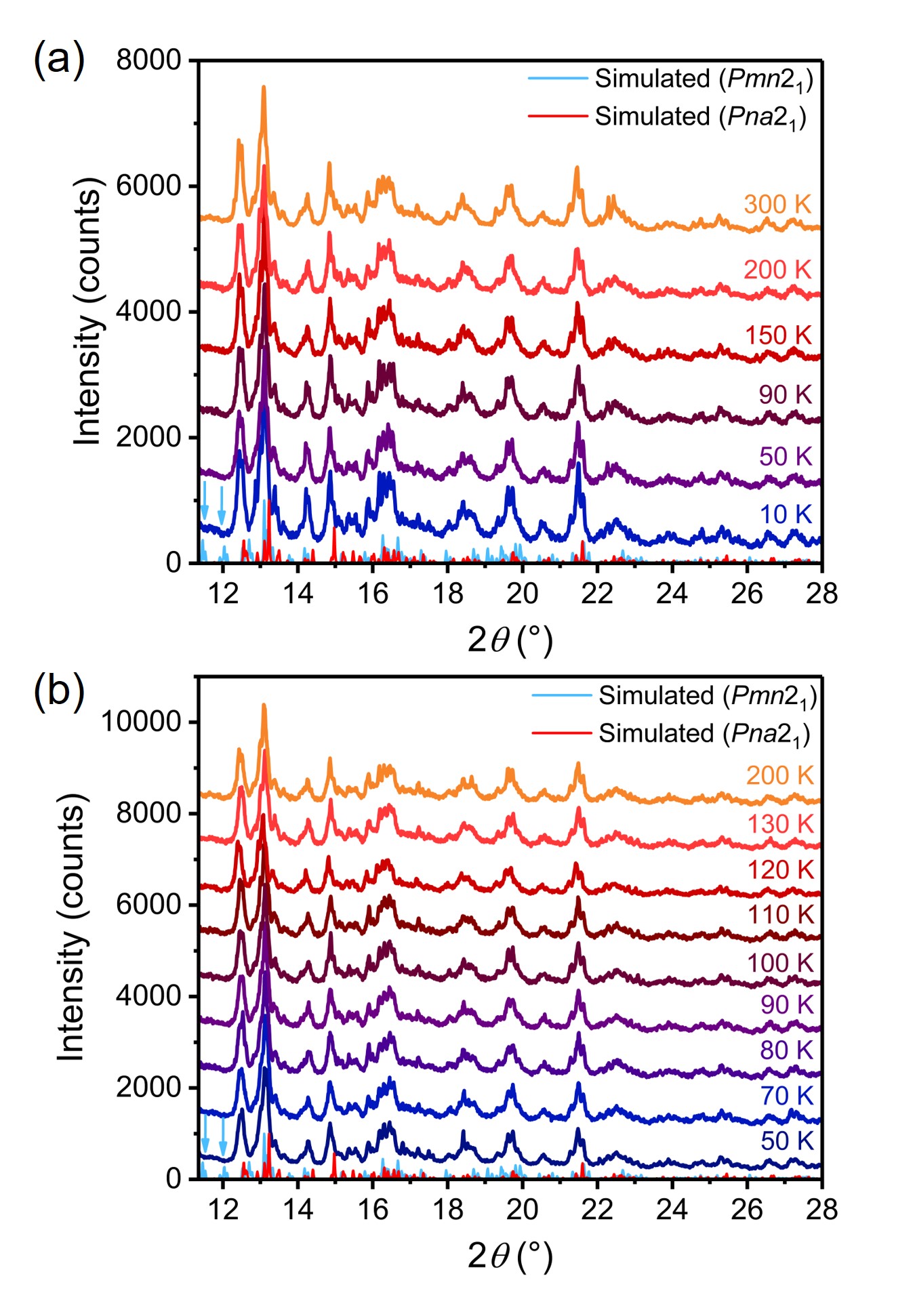}
    \caption{Variable temperature PXRD data for Ag$_8$SnS$_6$. The sample history is as follows: the data was first collected at 300 K; the sample was then quenched to 10 K and heated back to 200 K. The patterns were next obtained while cooling at temperatures between 150-50 K. (a) Patterns for the first warming sequence. (b) Subsequent patterns collected on cooling. All powder patterns are consistent with the rt Ag$_8$SnS$_6$ structure with no evidence for the lt phase. The light blue arrows mark the positions of several low angle peaks expected for the lt structure but not observed in any experimental patterns. The patterns are each offset by 1000 counts for clarity.}
\end{figure}

\clearpage
\newpage

\begin{figure}[!h]
    \centering
    \includegraphics[width=1\textwidth]{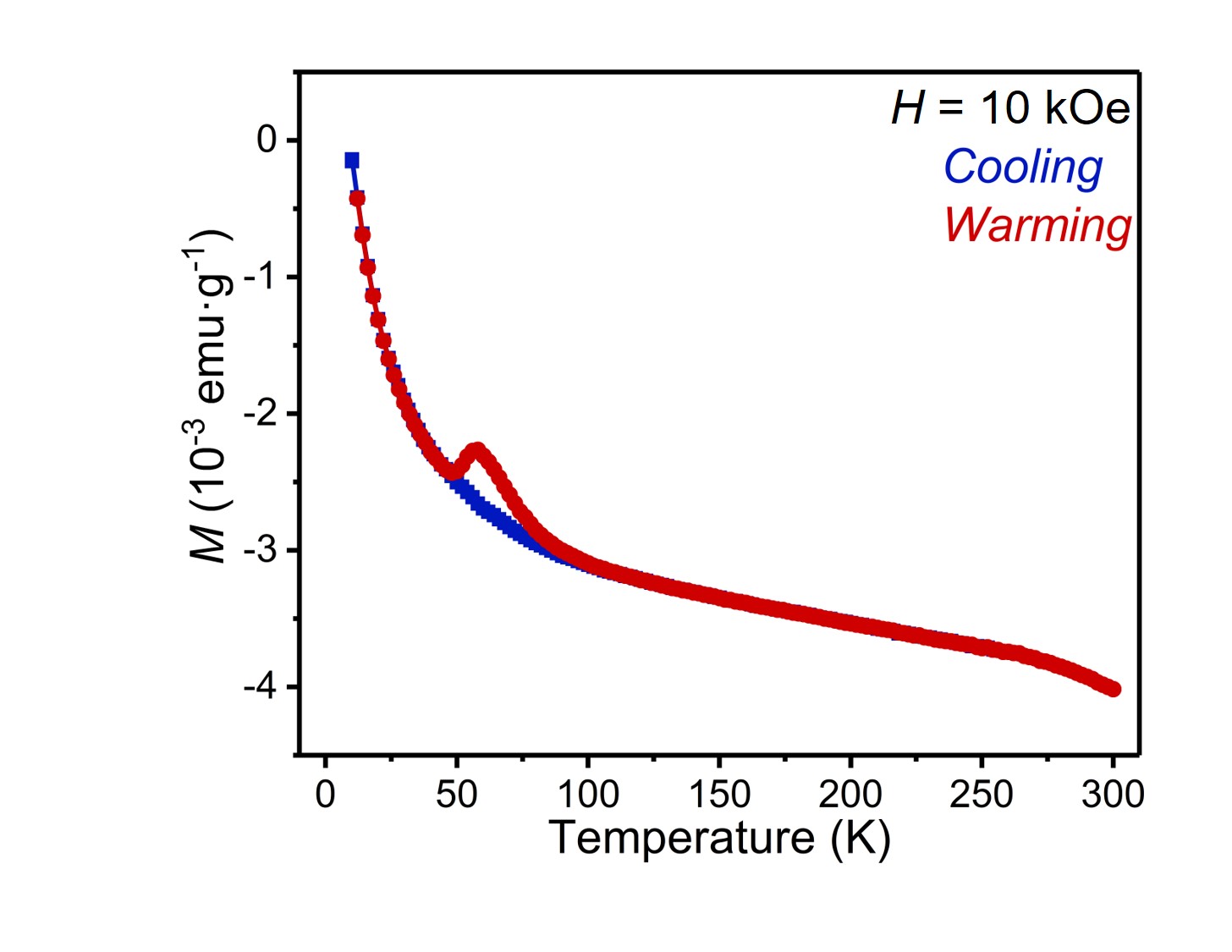}
    \caption{Temperature dependent magnetization for a powdered sample of Ag$_8$SnS$_6$.  The cooling dataset shows no evidence for the transition observed in the single crystals.  The feature above 50 K seen on warming is likely from trace oxygen.}
    \label{Figure S2}
\end{figure}

\clearpage
\newpage

\begin{figure}[!h]
    \centering
    \includegraphics[width=0.55\textwidth]{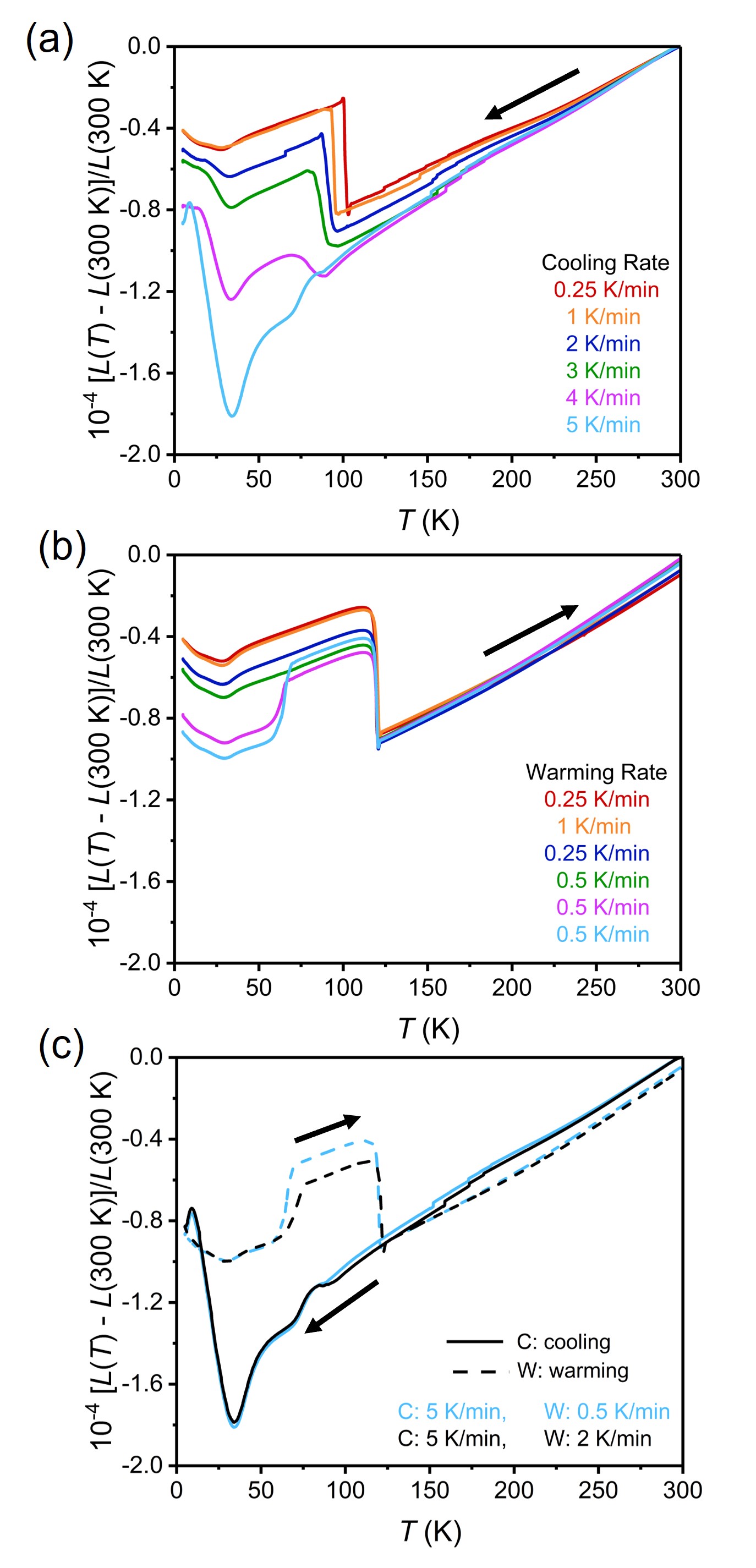}
    \caption{Thermal expansion for Ag$_8$SnS$_6$ measured at different cooling rates. (a) Data on cooling, (b) data on warming. (c) Data collected with: fast cooling and slow warming (black), and fast cooling and fast warming (light blue). The transition is steadily suppressed with increasing cooling rates, completely vanishing by 5 K/min. On warming, a transition into the lt structure is observed at $\approx$ 60 K for the measurements that were initially cooled faster than 3 K/min. The sample returns to the rt structure at 120 K on each warming dataset. The rapidly data collected after rapid cooling suggests the lt polymorph is recovered on warming at $\approx$ 60 K regardless of the heating rate. The feature at $\approx$ 30 K in each curve is an artifact from the strain gauge.}
    \label{Figure S4}
\end{figure}

\clearpage
\newpage

\begin{figure}[!h]
    \centering
    \includegraphics[width=1\textwidth]{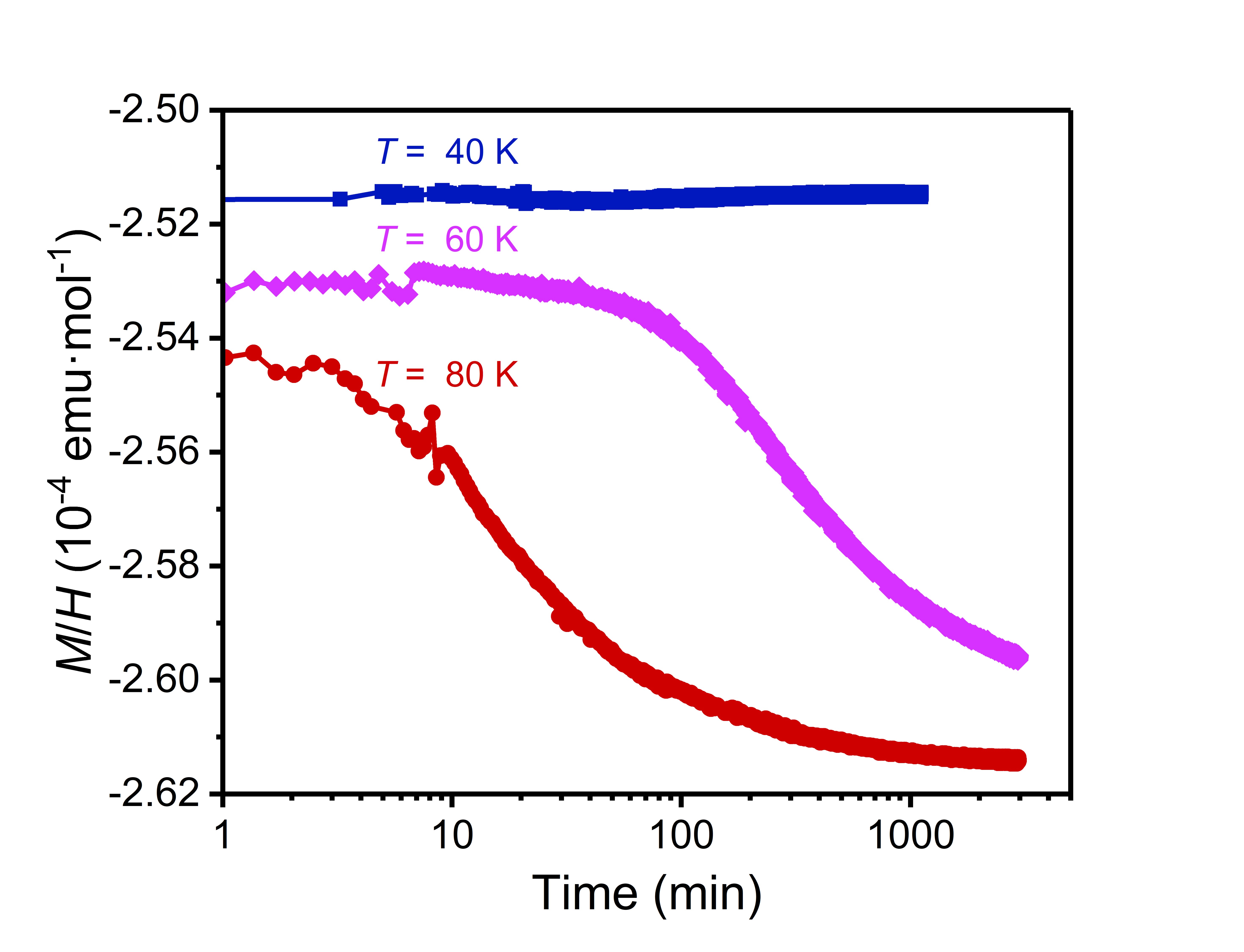}
    \caption{Magnetic susceptibility of a Ag$_8$SnS$_6$ single crystal as a function of time at \textit{T} = 40 K, 60 K, and 80 K. For each measurement, the sample was rapidly cooled ($\approx$ 12 K$\cdot$min$^{-1}$) from 300 K to the target temperature. The results suggest the sample begins relaxing to the lt polymorph within $\approx$ 10 minutes at 80 K and 100 minutes at 60 K, whereas at 40 K, the rt phase remains throughout the duration of the measurement. All measurements were conducted with \textit{H} = 10 kOe.} 
    \label{Figure S4}
\end{figure}

\clearpage
\newpage

\begin{figure}[!h]
    \centering
    \includegraphics[width=0.8\textwidth]{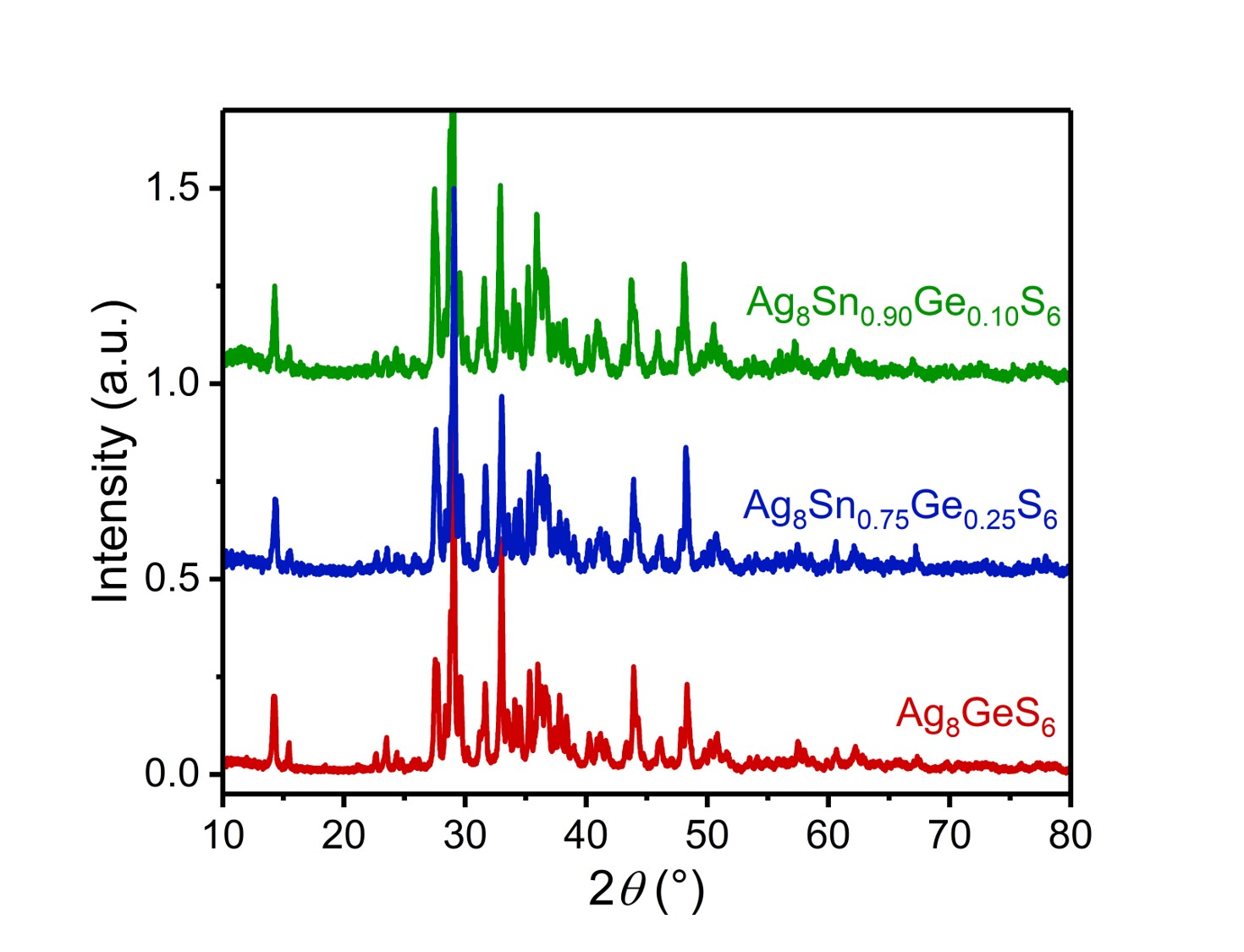}
    \caption{Powder XRD patterns for Ge-alloyed Ag$_8$Sn$_{1-x}$Ge$_x$S$_6$ and Ag$_8$GeS$_6$. All patterns are consistent with the room temperature \textit{Pna}2$_1$ structure of Ag$_8$GeS$_6$ and Ag$_8$SnS$_6$ with no discernible reflections from secondary phases and suggest successful Ge incorporation.}
    \label{Figure S4}
\end{figure}

\clearpage
\newpage

\begin{figure}[!h]
    \centering
    \includegraphics[width=1\textwidth]{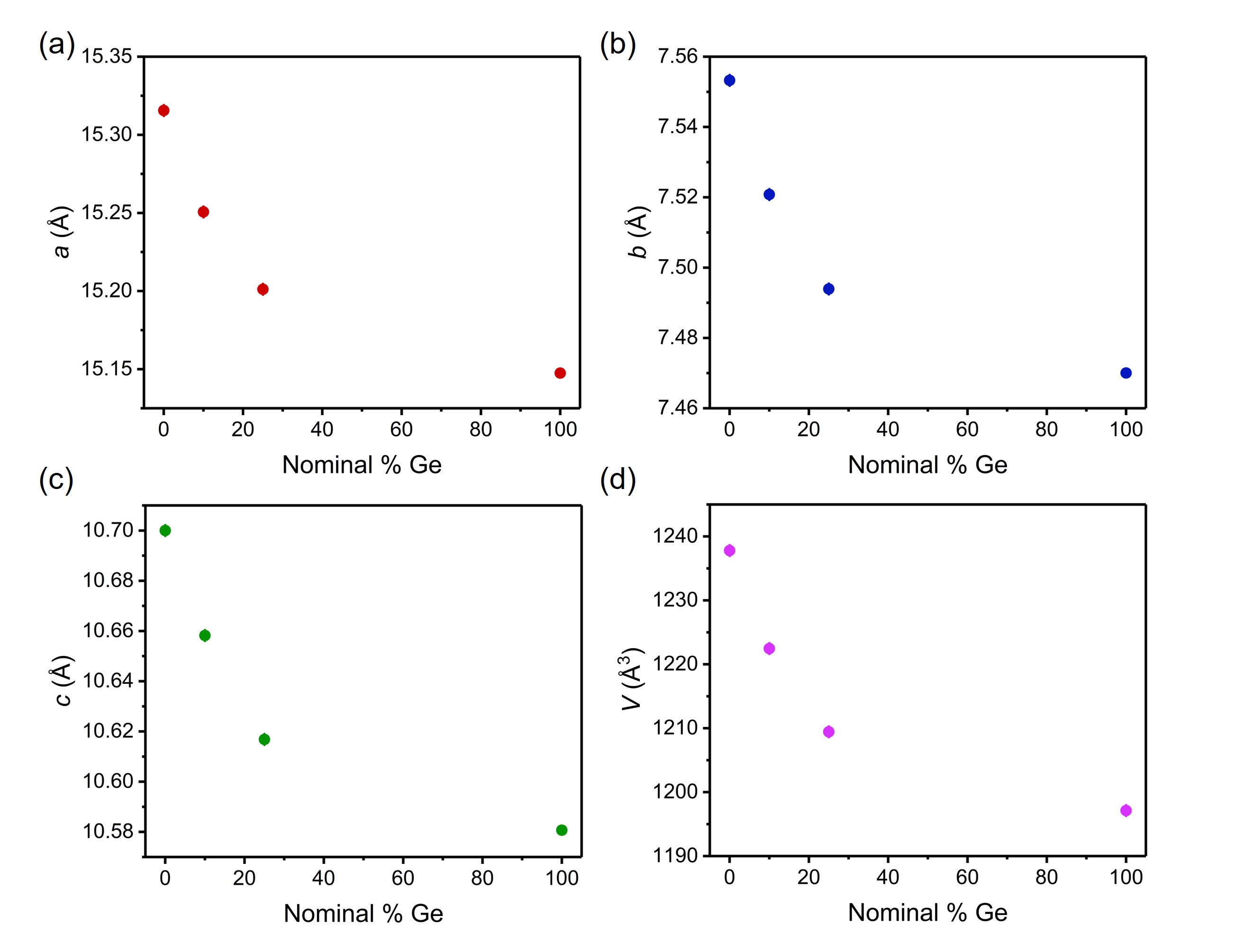}
    \caption{Refined lattice constants and unit cell volumes for Ge-alloyed Ag$_8$Sn$_{1-x}$Ge$_x$S$_6$ and pure Ag$_8$GeS$_6$.  The Ge fractions are based on the nominal Ge:Sn content used in the crystal growth.}
    \label{Figure S5}
\end{figure}

\clearpage
\newpage

\begin{figure}[!h]
    \centering
    \includegraphics[width=0.75\textwidth]{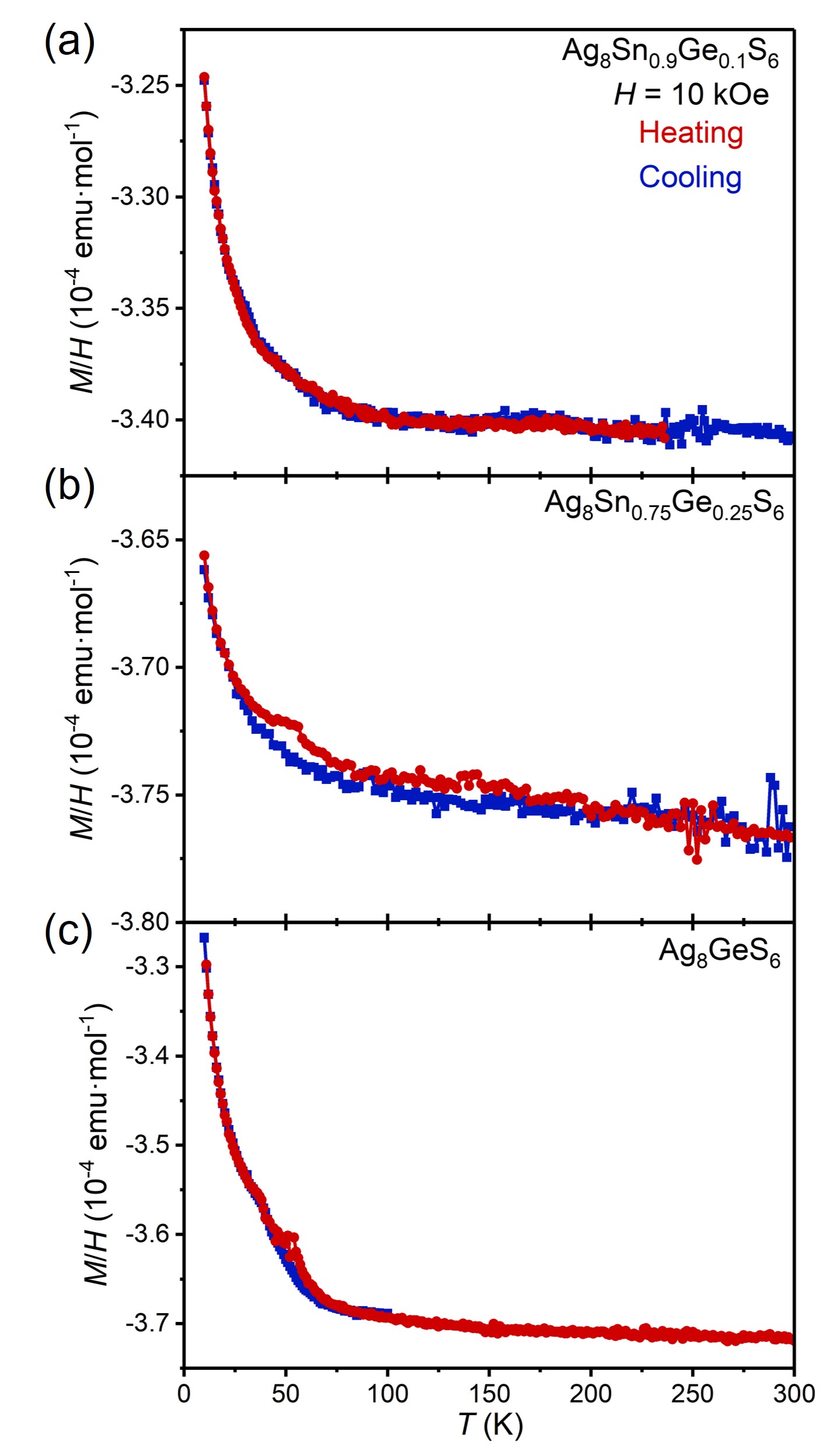}
    \caption{Temperature dependent magnetization data collected on single crystals of Ag$_8$GeS$_6$ and Ge-alloyed Ag$_8$Sn$_{1-x}$Ge$_x$S$_6$. The transition seen in pure Ag$_8$SnS$_6$ is not observed in Ge containing samples.}
    \label{Figure S6}
\end{figure}

\clearpage
\newpage

% Table generated by Excel2LaTeX from sheet 'Sheet6'
\begin{table}[htbp]
  \small
  \centering
  \caption{Single crystal data and structural refinement information for Ag$_8$SnS$_6$ at 295 K, 160 K, and 90 K.}
    \begin{tabular}{p{7.855em}llll}
    \toprule
    Phase & \multicolumn{3}{p{21.435em}}{rt Ag$_8$SnS$_6$} & \multicolumn{1}{p{8.07em}}{lt Ag$_8$SnS$_6$} \\
    \midrule
    Space group, \textit{Z} & \multicolumn{3}{p{21.435em}}{\textit{Pna}2$_1$, (\#33), 4} & \multicolumn{1}{p{8.07em}}{\textit{Pmn}2$_1$, (\#31), 2} \\
    \textit{a}, Å & \multicolumn{1}{p{7.145em}}{15.2993(9)} & \multicolumn{1}{p{7.145em}}{15.295(2)} & \multicolumn{1}{p{7.145em}}{15.2724(9)} & \multicolumn{1}{p{8.07em}}{7.6629(5)} \\
    \textit{b}, Å & \multicolumn{1}{p{7.145em}}{7.5479(4)} & \multicolumn{1}{p{7.145em}}{7.554(1)} & \multicolumn{1}{p{7.145em}}{7.5197(5)} & \multicolumn{1}{p{8.07em}}{7.5396(5)} \\
    \textit{c}, Å & \multicolumn{1}{p{7.145em}}{10.7045(6)} & \multicolumn{1}{p{7.145em}}{10.685(1)} & \multicolumn{1}{p{7.145em}}{10.6569(7)} & \multicolumn{1}{p{8.07em}}{10.6300(6)} \\
    \textit{V}, Å & \multicolumn{1}{p{7.145em}}{1236.1(1)} & \multicolumn{1}{p{7.145em}}{1234.5(2)} & \multicolumn{1}{p{7.145em}}{1223.9(1)} & \multicolumn{1}{p{8.07em}}{614.15(7)} \\
    $\rho$, g/cm$^3$ & 6.308 & 6.317 & 6.372 & 6.349 \\
    Crystal size, mm & \multicolumn{4}{p{29.505em}}{0.05 × 0.04× 0.02} \\
    Wavelength, Å  & \multicolumn{4}{p{29.505em}}{0.71073 (Mo \textit{K}$_\alpha$)} \\
    Temperature, K & 295 & 160 & 90  & 90 \\
    Abs Coef., mm$^{-1}$ & 15.332 & 15.353 & 15.486 & 15.43 \\
    Range of $\theta$, º & \multicolumn{1}{p{7.145em}}{2.66-27.24} & \multicolumn{1}{p{7.145em}}{2.66-27.45} & \multicolumn{1}{p{7.145em}}{2.67-27.53} & \multicolumn{1}{p{8.07em}}{2.70-27.51} \\
    Range of \textit{h}, \textit{k}, \textit{l} & \multicolumn{1}{p{7.145em}}{±19, ±9, -13 /+11} & \multicolumn{1}{p{7.145em}}{±19, ±9, -12 /+13} & \multicolumn{1}{p{7.145em}}{±19, ±9, -12 /+13} & \multicolumn{1}{p{8.07em}}{±9, ±9, -13 /+11} \\
    \textit{F}(000) & 2088 & 2088 & 2088 & 1044 \\
    Measured reflections & 21048 & 11425 & 19738 & 13402 \\
    Independent reflections & \multicolumn{1}{p{7.145em}}{2736 (\textit{R}$_{\text{int}}$ = 0.077)} & \multicolumn{1}{p{7.145em}}{2710 (\textit{R}$_{\text{int}}$ = 0.134)} & \multicolumn{1}{p{7.145em}}{2712 (\textit{R}$_{\text{int}}$ = 0.086)} & \multicolumn{1}{p{8.07em}}{1454 (\textit{R}$_{\text{int}}$ = 0.069)} \\
    Reflections with F $>$ 2$\sigma$(\textit{F}) & \multicolumn{1}{p{7.145em}}{2282 (\textit{R}$_{\text{$\sigma$}}$ = 0.051)} & \multicolumn{1}{p{7.145em}}{2055  (\textit{R}$_{\text{$\sigma$}}$ = 0.115)} & \multicolumn{1}{p{7.145em}}{2523 (\textit{R}$_{\text{$\sigma$}}$ = 0.053)} & \multicolumn{1}{p{8.07em}}{1400 (\textit{R}$_{\text{$\sigma$}}$ = 0.037)} \\
    GOF for \textit{F}2 & 1.12 & 1.04 & 1.1 & 1.21 \\
    Final R-indices [\textit{F} $>$ 2$\sigma$(\textit{F})] & \multicolumn{1}{p{7.145em}}{\textit{R}$_1$ = 0.048, \textit{wR}$_{\text{2}}$ = 0.110 } & \multicolumn{1}{p{7.145em}}{\textit{R}$_1$ = 0.055, \textit{wR}$_{\text{2}}$ = 0.113} & \multicolumn{1}{p{7.145em}}{\textit{R}$_1$ = 0.037, \textit{wR}$_{\text{2}}$ = 0.090} & \multicolumn{1}{p{8.07em}}{\textit{R}$_1$ = 0.031, \textit{wR}$_{\text{2}}$ = 0.076} \\
    \textit{R}-indices (all data) & \multicolumn{1}{p{7.145em}}{\textit{R}$_1$ = 0.067, \textit{wR}$_{\text{2}}$ = 0.118} & \multicolumn{1}{p{7.145em}}{\textit{R}$_1$ = 0.091, \textit{wR}$_{\text{2}}$ = 0.124} & \multicolumn{1}{p{7.145em}}{\textit{R}$_1$ = 0.042, \textit{wR}$_{\text{2}}$ = 0.092} & \multicolumn{1}{p{8.07em}}{\textit{R}$_1$ = 0.034, \textit{wR}$_{\text{2}}$ = 0.077} \\
    Larg. diff. peak and hole, e/Å$^{3}$ & \multicolumn{1}{p{7.145em}}{2.85 / -1.97} & \multicolumn{1}{p{7.145em}}{2.45 / -1.89} & \multicolumn{1}{p{7.145em}}{2.31 / -1.76} & \multicolumn{1}{p{8.07em}}{1.87 / -1.90} \\
    CSD number* & 2105238 & 2105240 & 2105241 & 2105242 \\
    \bottomrule
    \end{tabular}%
  \label{tab:addlabel}%
  \caption*{*Further details of the crystal structure refinement can be obtained from the Fachinformationszentrum Karlsruhe, 76344 Eggenstein-Leopoldshafen, Germany (fax: (+49)7247-808-666; e-mail: crysdata@fiz-karlsruhe.de) on quoting the depository number CSD 2105238, CSD 2105240, CSD 2105241, CSD 2105242.}
\end{table}%

\clearpage
\newpage

% Table generated by Excel2LaTeX from sheet 'rt-295K'
\begin{table}[!htbp]
  \small
  \centering
  \caption{Atomic positions and site occupancy for rt (\textit{Pna}2$_1$) Ag$_8$SnS$_6$ at 295 K.}
    \begin{tabular}{lllllr}
    \toprule
    Label & Wyckoff & x     & y     & z     & \multicolumn{1}{l}{Occupancy} \\
    \midrule
    Ag1   & 4a    & 0.01987(16) & 0.0168(3) & 0.0164(3) & 1 \\
    Ag2   & 4a    & 0.06054(16) & 0.2270(3) & 0.2486(2) & 1 \\
    Ag3   & 4a    & 0.12369(15) & 0.2177(4) & 0.7970(2) & 1 \\
    Ag4   & 4a    & 0.22444(15) & 0.0106(3) & 0.0072(3) & 1 \\
    Ag5   & 4a    & 0.25715(16) & 0.1518(4) & 0.3416(3) & 1 \\
    Ag6   & 4a    & 0.27508(14) & 0.3817(3) & 0.1023(2) & 1 \\
    Ag7   & 4a    & 0.41259(17) & 0.0807(3) & 0.1204(3) & 1 \\
    Ag8   & 4a    & 0.43476(16) & 0.0605(3) & 0.4315(2) & 1 \\
    Sn1   & 4a    & 0.12618(8) & 0.73105(16) & 0.26665(13) & 1 \\
    S1    & 4a    & 0.0003(3) & 0.2656(6) & 0.6380(5) & 1 \\
    S2    & 4a    & 0.1240(4) & 0.2701(7) & 0.0272(6) & 1 \\
    S3    & 4a    & 0.1224(3) & 0.4774(7) & 0.3991(5) & 1 \\
    S4    & 4a    & 0.2475(3) & 0.2352(7) & 0.6387(5) & 1 \\
    S5    & 4a    & 0.3871(3) & 0.3107(7) & 0.2806(6) & 1 \\
    S6    & 4a    & 0.6269(3) & 0.5162(7) & 0.3994(5) & 1 \\
    \bottomrule
    \end{tabular}%
  \label{tab:addlabel}%
\end{table}%

% Table generated by Excel2LaTeX from sheet 'rt-295K'
\begin{table}[!htbp]
  \small
  \centering
  \caption{Anisotropic displacement parameters for rt (\textit{Pna}2$_1$) Ag$_8$SnS$_6$ at 295 K.}
    \begin{tabular}{lllllll}
    \toprule
    Label & U11   & U22   & U33   & U23   & U13   & U12 \\
    \midrule
    Ag1   & 0.0485(14) & 0.0438(13) & 0.0701(18) & 0.0282(12) & -0.0320(13) & -0.0244(11) \\
    Ag2   & 0.0553(13) & 0.0403(11) & 0.0325(12) & 0.0014(10) & 0.0002(10) & 0.0025(10) \\
    Ag3   & 0.0351(12) & 0.107(2) & 0.0207(12) & 0.0008(12) & 0.0011(9) & 0.0162(13) \\
    Ag4   & 0.0375(12) & 0.0382(12) & 0.0757(18) & 0.0238(12) & 0.0182(12) & 0.0171(10) \\
    Ag5   & 0.0322(12) & 0.0715(18) & 0.088(2) & 0.0183(15) & 0.0055(13) & -0.0264(12) \\
    Ag6   & 0.0378(11) & 0.0376(11) & 0.0386(11) & -0.0055(9) & -0.0058(10) & 0.0047(9) \\
    Ag7   & 0.0595(15) & 0.0518(15) & 0.0468(14) & -0.0314(13) & -0.0076(12) & 0.0121(12) \\
    Ag8   & 0.0556(14) & 0.0333(11) & 0.0359(12) & 0.0086(10) & -0.0082(10) & -0.0070(10) \\
    Sn1   & 0.0129(6) & 0.0110(6) & 0.0103(7) & 0.0012(6) & 0.0002(5) & -0.0003(5) \\
    S1    & 0.015(3) & 0.017(3) & 0.016(3) & 0.000(2) & 0.0040(19) & -0.0018(19) \\
    S2    & 0.020(3) & 0.016(3) & 0.025(3) & -0.002(2) & 0.001(2) & 0.000(2) \\
    S3    & 0.017(2) & 0.011(2) & 0.018(3) & 0.006(2) & -0.001(2) & 0.0002(19) \\
    S4    & 0.017(3) & 0.017(3) & 0.013(3) & 0.000(2) & -0.0043(19) & -0.0021(19) \\
    S5    & 0.019(3) & 0.017(2) & 0.025(3) & -0.004(2) & 0.000(2) & 0.000(2) \\
    S6    & 0.017(3) & 0.017(2) & 0.016(3) & 0.004(2) & 0.000(2) & 0.002(2) \\
    \bottomrule
    \end{tabular}%
  \label{tab:addlabel}%
\end{table}%

\clearpage
\newpage

% Table generated by Excel2LaTeX from sheet 'lt-90K'
\begin{table}[htbp]
  \small
  \centering
  \caption{Atomic positions and site occupancy for lt (\textit{Pmn}2$_1$) Ag$_8$SnS$_6$ at 90 K.}
    \begin{tabular}{llrllr}
    \toprule
    Label & Wyckoff & \multicolumn{1}{l}{x} & y     & z     & \multicolumn{1}{l}{Occupancy} \\
    \midrule
    Sn    & 2a    & 0     & 0.7542(2) & 0.13475(17) & 1 \\
    Ag1   & 2a    & 0     & 0.2827(3) & 0.17454(19) & 1 \\
    Ag2   & 4b    & \multicolumn{1}{l}{0.20469(18)} & 0.46569(19) & 0.38326(14) & 1 \\
    Ag3   & 2a    & 0     & 0.3786(3) & 0.61464(19) & 1 \\
    Ag4   & 4b    & \multicolumn{1}{l}{0.2989(2)} & 0.10295(19) & 0.29177(13) & 1 \\
    Ag5   & 4b    & \multicolumn{1}{l}{0.3036(2)} & 0.1552(2) & 0.00045(16) & 1 \\
    S1    & 4b    & \multicolumn{1}{l}{0.2433(6)} & 0.2452(6) & 0.7608(4) & 1 \\
    S2    & 2a    & 0     & 0.1997(9) & 0.4034(6) & 1 \\
    S3    & 2a    & 0     & 0.5021(8) & 0.0010(7) & 1 \\
    S4    & 2a    & 0     & 0.0068(9) & 0.0000(7) & 1 \\
    S5    & 2a    & 0     & 0.7123(9) & 0.6555(7) & 1 \\
    \bottomrule
    \end{tabular}%
  \label{tab:addlabel}%
\end{table}%

% Table generated by Excel2LaTeX from sheet 'lt-90K'
\begin{table}[htbp]
  \small
  \centering
  \caption{Anisotropic displacement parameters for lt (\textit{Pmn}2$_1$) Ag$_8$SnS$_6$ at 90 K.}
    \begin{tabular}{lllllrr}
    \toprule
    Label & U11   & U22   & U33   & U23   & \multicolumn{1}{l}{U13} & \multicolumn{1}{l}{U12} \\
    \midrule
    Sn    & 0.0080(7) & 0.0030(8) & 0.0034(8) & -0.0004(6) & 0     & 0 \\
    Ag1   & 0.0130(9) & 0.0125(10) & 0.0040(9) & 0.0026(8) & 0     & 0 \\
    Ag2   & 0.0134(7) & 0.0095(8) & 0.0110(7) & 0.0003(6) & \multicolumn{1}{l}{-0.0007(6)} & \multicolumn{1}{l}{-0.0031(6)} \\
    Ag3   & 0.0131(10) & 0.0119(10) & 0.0091(11) & -0.0008(8) & 0     & 0 \\
    Ag4   & 0.0118(7) & 0.0119(7) & 0.0128(8) & 0.0022(6) & \multicolumn{1}{l}{0.0008(6)} & \multicolumn{1}{l}{-0.0012(6)} \\
    Ag5   & 0.0142(7) & 0.0213(8) & 0.0148(8) & -0.0028(7) & \multicolumn{1}{l}{-0.0042(7)} & \multicolumn{1}{l}{-0.0035(6)} \\
    S1    & 0.009(2) & 0.006(2) & 0.004(2) & -0.0001(18) & \multicolumn{1}{l}{0.0021(18)} & \multicolumn{1}{l}{-0.0001(17)} \\
    S2    & 0.011(3) & 0.009(3) & 0.005(3) & 0.001(2) & 0     & 0 \\
    S3    & 0.010(3) & 0.006(3) & 0.004(3) & -0.001(3) & 0     & 0 \\
    S4    & 0.011(3) & 0.007(3) & 0.004(3) & 0.000(3) & 0     & 0 \\
    S5    & 0.009(3) & 0.007(3) & 0.016(4) & 0.000(3) & 0     & 0 \\
    \bottomrule
    \end{tabular}%
  \label{tab:addlabel}%
\end{table}%

\clearpage
\newpage

% Table generated by Excel2LaTeX from sheet 'rt-90K'
\begin{table}[!htbp]
  \small
  \centering
  \caption{Atomic positions and site occupancy for rt (\textit{Pna}2$_1$) Ag$_8$SnS$_6$ at 90 K.}
    \begin{tabular}{lllllr}
    \toprule
    Label & Wyckoff & x     & y     & z     & \multicolumn{1}{l}{Occupancy} \\
    \midrule
    Ag1   & 4a    & 0.02027(9) & 0.0182(2) & 0.01868(16) & 1 \\
    Ag2   & 4a    & 0.05795(9) & 0.22828(19) & 0.24965(15) & 1 \\
    Ag3   & 4a    & 0.12349(10) & 0.2267(2) & 0.79698(17) & 1 \\
    Ag4   & 4a    & 0.22426(9) & 0.0110(2) & 0.00750(17) & 1 \\
    Ag5   & 4a    & 0.25610(9) & 0.1586(2) & 0.34069(17) & 1 \\
    Ag6   & 4a    & 0.27646(9) & 0.3770(2) & 0.10200(15) & 1 \\
    Ag7   & 4a    & 0.40981(10) & 0.0825(2) & 0.12448(16) & 1 \\
    Ag8   & 4a    & 0.43340(9) & 0.0622(2) & 0.42759(14) & 1 \\
    Sn1   & 4a    & 0.12664(7) & 0.72938(17) & 0.26581(13) & 1 \\
    S1    & 4a    & 0.0000(3) & 0.2691(6) & 0.6363(5) & 1 \\
    S2    & 4a    & 0.1243(3) & 0.2745(7) & 0.0294(5) & 1 \\
    S3    & 4a    & 0.1213(3) & 0.4775(6) & 0.4000(5) & 1 \\
    S4    & 4a    & 0.2472(3) & 0.2314(6) & 0.6369(5) & 1 \\
    S5    & 4a    & 0.3872(3) & 0.3216(6) & 0.2826(5) & 1 \\
    S6    & 4a    & 0.6271(3) & 0.5154(6) & 0.3991(5) & 1 \\
    \bottomrule
    \end{tabular}%
  \label{tab:addlabel}%
\end{table}%

% Table generated by Excel2LaTeX from sheet 'rt-90K'
\begin{table}[!htbp]
  \small
  \centering
  \caption{Anisotropic displacement parameters for rt (\textit{Pna}2$_1$) Ag$_8$SnS$_6$ at 90 K.}
    \begin{tabular}{lllllll}
    \toprule
    Label & U11   & U22   & U33   & U23   & U13   & U12 \\
    \midrule
    Ag1   & 0.0165(7) & 0.0137(8) & 0.0197(8) & 0.0059(7) & -0.0076(7) & -0.0056(6) \\
    Ag2   & 0.0154(7) & 0.0130(7) & 0.0108(7) & 0.0001(6) & 0.0008(6) & 0.0008(6) \\
    Ag3   & 0.0149(8) & 0.0294(10) & 0.0067(8) & -0.0017(7) & 0.0000(5) & 0.0053(6) \\
    Ag4   & 0.0127(6) & 0.0116(7) & 0.0256(9) & 0.0069(7) & 0.0053(6) & 0.0045(6) \\
    Ag5   & 0.0118(6) & 0.0224(9) & 0.0225(8) & 0.0059(7) & 0.0006(6) & -0.0065(6) \\
    Ag6   & 0.0131(6) & 0.0124(7) & 0.0120(7) & -0.0013(6) & -0.0021(6) & 0.0016(6) \\
    Ag7   & 0.0179(7) & 0.0168(9) & 0.0164(8) & -0.0094(7) & -0.0032(6) & 0.0033(6) \\
    Ag8   & 0.0176(7) & 0.0124(8) & 0.0116(8) & 0.0025(6) & -0.0017(6) & -0.0023(6) \\
    Sn1   & 0.0059(6) & 0.0075(6) & 0.0032(7) & 0.0008(5) & -0.0003(4) & 0.0000(4) \\
    S1    & 0.0069(19) & 0.008(2) & 0.004(2) & -0.0026(17) & 0.0033(15) & -0.0018(16) \\
    S2    & 0.009(2) & 0.010(2) & 0.009(3) & 0.0011(18) & 0.0008(16) & -0.0028(16) \\
    S3    & 0.009(2) & 0.007(2) & 0.006(2) & 0.0030(18) & 0.0009(16) & -0.0008(16) \\
    S4    & 0.007(2) & 0.009(2) & 0.009(2) & 0.0004(17) & -0.0028(16) & 0.0004(15) \\
    S5    & 0.008(2) & 0.008(2) & 0.012(2) & 0.001(2) & -0.0006(17) & -0.0023(17) \\
    S6    & 0.010(2) & 0.008(2) & 0.006(2) & -0.0001(18) & 0.0009(16) & 0.0020(16) \\
    \bottomrule
    \end{tabular}%
  \label{tab:addlabel}%
\end{table}%

\clearpage
\newpage

% Table generated by Excel2LaTeX from sheet 'rt-160K'
\begin{table}[!htbp]
  \small
  \centering
  \caption{Atomic positions and site occupancy for rt (\textit{Pna}2$_1$) Ag$_8$SnS$_6$ at 160 K.}
    \begin{tabular}{lllllr}
    \toprule
    Label & Wyckoff & x     & y     & z     & \multicolumn{1}{l}{Occupancy} \\
    \midrule
    Ag1   & 4a    & 0.02008(16) & 0.0178(4) & 0.0172(3) & 1 \\
    Ag2   & 4a    & 0.05870(17) & 0.2274(3) & 0.2493(2) & 1 \\
    Ag3   & 4a    & 0.12383(17) & 0.2246(4) & 0.7971(2) & 1 \\
    Ag4   & 4a    & 0.22439(16) & 0.0107(4) & 0.0075(3) & 1 \\
    Ag5   & 4a    & 0.25636(17) & 0.1561(4) & 0.3408(3) & 1 \\
    Ag6   & 4a    & 0.27600(15) & 0.3789(4) & 0.1025(2) & 1 \\
    Ag7   & 4a    & 0.41079(18) & 0.0824(4) & 0.1229(3) & 1 \\
    Ag8   & 4a    & 0.43404(17) & 0.0619(4) & 0.4292(2) & 1 \\
    Sn1   & 4a    & 0.12646(12) & 0.7302(3) & 0.26620(17) & 1 \\
    S1    & 4a    & 0.0000(5) & 0.2660(10) & 0.6366(7) & 1 \\
    S2    & 4a    & 0.1242(5) & 0.2713(10) & 0.0287(7) & 1 \\
    S3    & 4a    & 0.1221(5) & 0.4767(11) & 0.3998(7) & 1 \\
    S4    & 4a    & 0.2472(5) & 0.2330(10) & 0.6376(6) & 1 \\
    S5    & 4a    & 0.3878(5) & 0.3183(11) & 0.2820(8) & 1 \\
    S6    & 4a    & 0.6277(5) & 0.5140(11) & 0.3990(7) & 1 \\
    \toprule
    \end{tabular}%
  \label{tab:addlabel}%
\end{table}%

% Table generated by Excel2LaTeX from sheet 'rt-160K'
\begin{table}[!htbp]
  \small
  \centering
  \caption{Anisotropic displacement parameters for rt (\textit{Pna}2$_1$) Ag$_8$SnS$_6$ at 160 K.}
    \begin{tabular}{lllllll}
    \toprule
    Label & U11   & U22   & U33   & U23   & U13   & U12 \\
    \midrule
    Ag1   & 0.0238(13) & 0.0214(17) & 0.0337(16) & 0.0125(13) & -0.0150(13) & -0.0117(12) \\
    Ag2   & 0.0236(13) & 0.0223(16) & 0.0152(13) & 0.0017(11) & 0.0008(11) & 0.0010(11) \\
    Ag3   & 0.0188(13) & 0.051(2) & 0.0098(13) & -0.0007(12) & 0.0005(10) & 0.0078(14) \\
    Ag4   & 0.0173(12) & 0.0184(16) & 0.0405(16) & 0.0126(13) & 0.0091(12) & 0.0078(12) \\
    Ag5   & 0.0153(12) & 0.036(2) & 0.0411(16) & 0.0094(14) & 0.0005(12) & -0.0126(13) \\
    Ag6   & 0.0181(11) & 0.0195(15) & 0.0191(12) & -0.0037(11) & -0.0030(11) & 0.0031(11) \\
    Ag7   & 0.0276(13) & 0.0277(19) & 0.0244(14) & -0.0184(13) & -0.0047(12) & 0.0054(13) \\
    Ag8   & 0.0277(13) & 0.0175(16) & 0.0183(13) & 0.0046(11) & -0.0036(11) & -0.0036(11) \\
    Sn1   & 0.0066(9) & 0.0052(12) & 0.0038(10) & 0.0013(9) & -0.0003(8) & 0.0002(8) \\
    S1    & 0.010(3) & 0.009(4) & 0.010(4) & -0.002(3) & 0.001(3) & -0.004(3) \\
    S2    & 0.014(3) & 0.003(4) & 0.012(4) & 0.000(3) & 0.000(3) & 0.003(3) \\
    S3    & 0.010(3) & 0.009(5) & 0.010(4) & 0.006(3) & 0.001(3) & -0.001(3) \\
    S4    & 0.009(3) & 0.010(4) & 0.003(3) & 0.000(3) & -0.003(3) & -0.001(3) \\
    S5    & 0.010(4) & 0.011(5) & 0.021(4) & -0.003(4) & -0.001(3) & 0.001(3) \\
    S6    & 0.010(3) & 0.007(4) & 0.012(4) & 0.003(3) & 0.001(3) & 0.001(3) \\
    \bottomrule
    \end{tabular}%
  \label{tab:addlabel}%
\end{table}%

\clearpage
\newpage

\begin{table}[!ht]
  \centering
  \caption{Unit cell parameters of Ag$_8$SnS$_6$ determined from single crystal X-ray diffraction analysis at 90 K, 160 K, 295 K.}
    \begin{tabular}{lllll}
    \toprule
    Unit cell, \textit{T} & lt (\textit{Pmn}2$_1$), 90 K & rt (\textit{Pna}2$_1$), 90 K & rt (\textit{Pna}2$_1$), 160 K & rt (\textit{Pna}2$_1$), 295 K \\
    \midrule
    \textit{a}, Å  & 7.6629(5) & 15.2724(9) & 15.295(2) & 15.2993(9) \\
    \textit{b}, Å  & 7.5396(5) & 7.5197(5) & 7.554(1) & 7.5479(4) \\
    \textit{c}, Å  & 10.6300(6) & 10.6569(7) & 10.685(1) & 10.7045(6) \\
    \textit{V}, Å$^3$ & 614.15(7) & 1223.9(1) & 1234.5(2) & 1236.1(1) \\
    \textit{V/Z}, Å$^3$ & 307.08(4) & 305.98(3) & 308.63(5) & 309.03(3) \\
    \bottomrule
    \end{tabular}%
  \label{tab:addlabel}%
\end{table}%

\clearpage
\newpage

\begin{table}[htbp]
  \centering
  \caption{Sn-S and Ag-S bond lengths for Ag$_8$SnS$_6$ determined from single crystal X-ray diffraction analysis at 90 K, 160 K, 295 K. The 90 K data for the rt \textit{Pna}2$_1$ was obtained by rapidly cooling the sample to 90 K, where the large thermal hysteresis of the transformation allowed us to collect data in the rt phase.}
    \begin{tabular}{rrrrllrll}
    \toprule
    \multicolumn{3}{c}{lt Ag$_8$SnS$_6$, 90 K} & \multicolumn{3}{c}{rt Ag$_8$SnS$_6$, 90 K} & \multicolumn{3}{c}{rt Ag$_8$SnS$_6$, 295K} \\
    \multicolumn{3}{c}{(Ag$_8$GeSe$_6$-type, \textit{Pmn}2$_1$)} & \multicolumn{3}{c}{(Ag$_8$GeS$_6$-type, \textit{Pna}2$_1$)} & \multicolumn{3}{c}{(Ag$_8$GeS$_6$-type, \textit{Pna}2$_1$)} \\
    \midrule
    \multicolumn{1}{l}{Sn1 (2a)} & \multicolumn{1}{l}{S3} & \multicolumn{1}{l}{2.373(7)} & \multicolumn{1}{l}{Sn1  (4a)} & S4    & 2.366(4) & \multicolumn{1}{l}{Sn1  (4a)} & S4    & 2.369(6) \\
          & \multicolumn{1}{l}{S1 } & \multicolumn{1}{l}{2.380(4)} &       & S3    & 2.375(5) &       & S3    & 2.376(6) \\
          & \multicolumn{1}{l}{S1} & \multicolumn{1}{l}{2.380(4)} &       & S1    & 2.376(4) &       & S1    & 2.377(6) \\
          & \multicolumn{1}{l}{S4 } & \multicolumn{1}{l}{2.383(7)} &       & S6    & 2.388(5) &       & S6    & 2.379(6) \\
    \multicolumn{1}{l}{Ag1  (2a)} & \multicolumn{1}{l}{S3} & \multicolumn{1}{l}{2.478(7)} & \multicolumn{1}{l}{Ag1 (4a)} & S2    & 2.501(5) & \multicolumn{1}{l}{Ag1 (4a)} & S2    & 2.491(7) \\
          & \multicolumn{1}{l}{S2} & \multicolumn{1}{l}{2.512(7)} &       & S1    & 2.517(4) &       & S1    & 2.513(6) \\
          & \multicolumn{1}{l}{S4} & \multicolumn{1}{l}{2.788(7)} &       & S6    & 2.586(4) &       & S6    & 2.569(6) \\
    \multicolumn{1}{l}{Ag2  (4b)} & \multicolumn{1}{l}{S2} & \multicolumn{1}{l}{2.555(6)} & \multicolumn{1}{l}{Ag2  (4a)} & S2    & 2.580(5) & \multicolumn{1}{l}{Ag2  (4a)} & S2    & 2.582(7) \\
          & \multicolumn{1}{l}{S1} & \multicolumn{1}{l}{2.570(5)} &       & S6    & 2.647(5) &       & S6    & 2.640(6) \\
          & \multicolumn{1}{l}{S3} & \multicolumn{1}{l}{2.597(4)} &       & S3    & 2.649(5) &       & S3    & 2.651(6) \\
    \multicolumn{1}{l}{Ag3 (2a)} & \multicolumn{1}{l}{S5} & \multicolumn{1}{l}{2.553(7)} &       & S5    & 2.658(5) &       & S5    & 2.694 (6) \\
          & \multicolumn{1}{l}{S2} & \multicolumn{1}{l}{2.619(7)} & \multicolumn{1}{l}{Ag3 (4a)} & S2    & 2.503(5) & \multicolumn{1}{l}{Ag3 (4a)} & S2    & 2.499(7) \\
          & \multicolumn{1}{l}{S1} & \multicolumn{1}{l}{2.627(5)} &       & S4    & 2.546(5) &       & S4    & 2.553(6) \\
          & \multicolumn{1}{l}{S1} & \multicolumn{1}{l}{2.627(5)} &       & S1    & 2.568(5) &       & S1    & 2.570(6) \\
    \multicolumn{1}{l}{Ag4  (4b)} & \multicolumn{1}{l}{S5} & \multicolumn{1}{l}{2.532(6)} & \multicolumn{1}{l}{Ag4  (4a)} & S2    & 2.512(5) & \multicolumn{1}{l}{Ag4  (4a)} & S2    & 2.496(7) \\
          & \multicolumn{1}{l}{S1} & \multicolumn{1}{l}{2.665(5)} &       & S4    & 2.552(5) &       & S4    & 2.541(6) \\
          & \multicolumn{1}{l}{S2} & \multicolumn{1}{l}{2.681(4)} &       & S3    & 2.634(4) &       & S3    & 2.627(6) \\
          & \multicolumn{1}{l}{S4} & \multicolumn{1}{l}{2.821(6)} & \multicolumn{1}{l}{Ag5  (4a)} & S5    & 2.427(5) & \multicolumn{1}{l}{Ag5  (4a)} & S5    & 2.413(6) \\
    \multicolumn{1}{l}{Ag5  (4b)} & \multicolumn{1}{l}{S5} & \multicolumn{1}{l}{2.446(6)} &       & S6    & 2.446(5) &       & S6    & 2.441(6) \\
          & \multicolumn{1}{l}{S4} & \multicolumn{1}{l}{2.582(3)} & \multicolumn{1}{l}{Ag6  (4a)} & S2    & 2.567(5) & \multicolumn{1}{l}{Ag6  (4a)} & S2    & 2.587(7) \\
          & \multicolumn{1}{l}{S1} & \multicolumn{1}{l}{2.676(5)} &       & S5    & 2.595(5) &       & S5    & 2.619(7) \\
          &       &       &       & S4    & 2.715(5) &       & S4    & 2.720(6) \\
          &       &       &       & S6    & 2.739(5) &       & S6    & 2.753(6) \\
          &       &       & \multicolumn{1}{l}{Ag7 (4a)} & S5    & 2.488(5) & \multicolumn{1}{l}{Ag7 (4a)} & S5    & 2.477(7) \\
          &       &       &       & S3    & 2.563(5) &       & S3    & 2.559(7) \\
          &       &       &       & S1    & 2.733(5) &       & S1    & 2.734(6) \\
          &       &       & \multicolumn{1}{l}{Ag8  (4a)} & S2    & 2.576(5) & \multicolumn{1}{l}{Ag8  (4a)} & S2    & 2.583(7) \\
          &       &       &       & S5    & 2.586(5) &       & S5    & 2.596(7) \\
          &       &       &       & S1    & 2.755(5) &       & S1    & 2.756(6) \\
          &       &       &       & S3    & 2.900(4) &       & S3    & 2.909(6) \\
          \bottomrule
    \end{tabular}%
  \label{tab:addlabel}%
\end{table}%

\end{doublespace}